\documentclass[12pt]{article}
\pdfoutput=1

\usepackage{amssymb}
\usepackage{color}
\usepackage{graphicx,psfrag}
\usepackage{amsmath}
\usepackage{hyperref}
\usepackage{bbold}

\usepackage[applemac]{inputenc}

\numberwithin{equation}{section}

\def\be{\begin{equation}}
\def\ee{\end{equation}}
\def\bea{\begin{eqnarray}}
\def\eea{\end{eqnarray}}
\def\bb{\bar{b}}
\def\ba{\bar{a}}
\def\bc{\bar{c}}

\def\bF{\bar{F}}

\def\bW{\overline{W}}
\def\bZ{\bar Z}
\def\bU{\overline U}
\def\bd{\bar d}
\def\M{M_{Pl}}

\def\hF{\hat{F}}

\begin{document}

\begin{titlepage}

\setcounter{page}{1} \baselineskip=15.5pt \thispagestyle{empty}

\bigskip\
\begin{center}
{\Large \bf  The Wasteland of Random Supergravities}
\vskip 5pt
\vskip 15pt
\end{center}
\vspace{0.5cm}
\begin{center}
{
\large
David Marsh, Liam McAllister, and Timm Wrase}
\end{center}

\vspace{0.1cm}

\begin{center}
\vskip 4pt
\textsl{Department of Physics, Cornell University,
Ithaca, NY 14853}

\end{center} %\vfil
{\small  \noindent  \\[0.2cm]
\noindent

We show that in a general
${\cal{N}}=1$ supergravity with $N \gg 1$ scalar fields, an exponentially small fraction of the de Sitter critical points are metastable vacua.
Taking the superpotential and K\"ahler potential to be random functions, we construct a random matrix model for the Hessian matrix, which is well-approximated by the sum of a Wigner matrix and two Wishart matrices.
We compute the eigenvalue spectrum analytically from the free convolution of the constituent spectra and find that in typical configurations, a significant fraction of the eigenvalues are negative.  Building on the Tracy-Widom law governing fluctuations of extreme eigenvalues,
we determine the probability $P$ of a large fluctuation in which all the eigenvalues become positive.
Strong eigenvalue repulsion makes this extremely unlikely: we find $P \propto {\rm exp}(-c\,N^p)$, with $c,\ p$ being constants. For generic critical points we find $p \approx 1.5$, while for approximately-supersymmetric critical points, $p\approx 1.3$.
Our results  have significant implications for
the counting of de Sitter vacua in string theory, but
the number of vacua remains vast.
}

\vspace{0.3cm}

\vspace{0.6cm}

\vfil
\begin{flushleft}
\small \today
\end{flushleft}
\end{titlepage}

\newpage
\tableofcontents
\newpage

\section{Introduction}
Perhaps the most pressing question in string theory is whether the theory admits solutions consistent with all observations.  In light of the discovery of the acceleration of the universe, it is essential to pursue de Sitter solutions of string theory, and to understand whether these solutions are so numerous that they can account for the smallness of the vacuum energy.  After a decade marked by significant advances in understanding flux compactifications \cite{Douglas:2006es}, there is now compelling, but still largely indirect, evidence for the existence of a vast landscape of metastable de Sitter vacua.    Direct enumeration of explicit de Sitter vacua remains a distant goal.

The cardinal difficulty in constructing de Sitter solutions is that in the absence of supersymmetry, the scalar potential can have instabilities along one or more directions in the scalar field space.  When the number of fields is large -- which is both generic in Calabi-Yau flux compactifications, and indispensable for providing an astronomical number of vacua -- direct examination of the Hessian matrix of the scalar potential becomes impractical.  This impasse motivates a statistical approach, in which the compactification data are taken to be random variables.

As a metastable vacuum is
a critical point of the scalar potential at which the Hessian matrix is positive definite, it is natural first to ascertain the statistical properties of general critical points, and then to characterize the subset of critical points that are in fact metastable vacua.  In the seminal work \cite{DD}, Denef and Douglas formulated this problem for a general four-dimensional ${\cal{N}}=1$ supergravity theory, taking the superpotential $W$ and K\"ahler potential $K$ to be random functions, in a precise sense that we shall review.   Denef and Douglas studied the eigenvalues of the Hessian matrix ${\cal{H}}$ and argued that a significant fraction of critical points are metastable vacua.

In this work we reexamine the stability of de Sitter critical points in a general four-dimensional ${\cal{N}}=1$ supergravity.  Our tool is random matrix theory:
${\cal{H}}$
is a large matrix, and a great deal can be said about its eigenvalue spectrum.
Moreover, given an ensemble of random matrices that typically have
negative eigenvalues, the Tracy-Widom theory of fluctuations of extreme eigenvalues allows one to compute the probability of drawing a positive-definite matrix from the ensemble \cite{TW}.  The key phenomenon is eigenvalue repulsion: a large fluctuation through which all eigenvalues become positive generally requires an increase in the local eigenvalue density, which is statistically costly.

We obtain results that depend on the relative sizes of the soft supersymmetry-breaking masses and the supersymmetric masses.   At a generic critical point, the supersymmetric masses are not hierarchically larger than the soft masses,
and supersymmetry provides negligible protection from instability.  We develop a random matrix model for ${\cal{H}}$ and obtain an analytic expression for its eigenvalue spectrum.
The
spectrum has considerable support at negative values, so that tachyons are generic.
Building on extensions of the Tracy-Widom theory due to Dean and Majumdar \cite{Dean2,Dean}, we then argue that the probability $P$ of a fluctuation rendering ${\cal{H}}$
positive-definite is
\begin{equation} \label{eq:prob}
 P \propto {\rm exp}(-c\, N^p) \,,
\end{equation}
at leading order in large $N$, where $N$ is the number of complex scalar fields, and $c$, $p$ are constants, with $p \approx 1.5$.
This is qualitatively similar to the original results of Aazami and Easther \cite{Aazami:2005jf},
who obtained $p \approx 2$ in a simpler model of the Hessian matrix.
We conclude that an exceedingly small fraction of generic critical points are  metastable.

The more promising regime, as stressed by Denef and Douglas, is that in which approximate supersymmetry protects against instabilities.  When the soft masses are small compared to the supersymmetric masses, the only significant risk of an instability arises from the direction parameterized by the scalar superpartner of the
Goldstino,\footnote{For extensive investigations of this unstable direction, see \cite{Covi}.} and we show that at a critical point that is generic apart from this requirement of approximate supersymmetry, there are almost always two tachyons.  We identify a negative-definite contribution to these eigenvalues that is a manifestation of eigenvalue repulsion between the Goldstino directions and the supersymmetrically-stabilized scalars.
The eigenvalue repulsion contribution is dominant at large $N$, so that it is extremely improbable that both eigenvalues fluctuate to become positive:
the probability of positivity
again takes the form (\ref{eq:prob}),
but now with $p \approx 1.3$.

Although our results clearly demonstrate that an exponentially small fraction of the critical points of a generic random supergravity theory are metastable, this finding in no way precludes the existence of a landscape of metastable de Sitter vacua.  First of all, the actual number of metastable vacua can be extremely large -- and in particular, larger than $10^{120}$ --  while still being exponentially small compared to the number of critical points.  Second, our analysis applies when $K$ and $W$ are generic functions of all of their arguments.  A well-motivated configuration violating this assumption is a theory involving two decoupled sectors, containing $N_H$ and $N_L$ scalar fields,
respectively.   If the $N_H$ fields receive large supersymmetric masses and dynamical supersymmetry breaking occurs at a lower scale in the sector containing $N_L$ fields, then the dominant factor in the number of critical points can be  exponential in $N_H$, while the fraction of critical points that are unstable is proportional to ${\rm exp}(-c N_L^p)$.  Thus, for $N_H \gg N_L$ our findings yield only a modest reduction in the number of vacua.

The organization of this paper is as follows.  In \S\ref{sec:cp} we set our notation and review the structure of the Hessian matrix ${\cal{H}}$ at a critical point in a general four-dimensional ${\cal N}=1$ supergravity, following \cite{DD}.  In \S\ref{sec:RMT} we introduce the ideas from random matrix theory that are essential in this work, reviewing the relevant ensembles and assembling results from the theory of fluctuations of extreme eigenvalues.
In \S\ref{sec:MMW} we apply these methods to study ${\cal{H}}$
at a generic critical point.
We compute the eigenvalue spectrum analytically and obtain the probability of a large fluctuation that renders ${\cal{H}}$
positive-definite.  In \S\ref{sec:DD} we study ${\cal{H}}$
at an approximately-supersymmetric critical point.  We show that eigenvalue repulsion typically leads to two negative eigenvalues associated to the Goldstino, and we again compute the probability of a fluctuation to positivity.  In \S\ref{sec:KKLT} we discuss extensions of our assumptions, and we illustrate our results in the KKLT scenario.  We conclude in \S\ref{sec:conclusions}.

\section{Critical Points in ${\cal{N}}=1$ Supergravity}
\label{sec:cp}
In this section we will discuss the form of the critical point equation, and describe the structure of
${\cal{H}}$ at a critical point, in a general four-dimensional ${\cal N}=1$ supergravity.
In \S\ref{sec:cpM} and \S\ref{sec:Hess} we closely follow the work of Denef and Douglas \cite{DD}, reviewing how the critical point equation can be written as an eigenvalue equation,
and we establish notation for the different contributions to the Hessian matrix.
Then, in \S\ref{sec:RS}, we give a precise definition of a random supergravity, whose critical points will be the object of study
in \S\ref{sec:MMW} and \S\ref{sec:DD}.

The
F-term
potential in  an ${\cal N} =1$ supergravity with $N$ chiral superfields is given by
 \bea
 V = e^{K} \left( F_{a} \bar{F}^{a} - 3 |W|^2 \right) \, ,
 \eea with $a=1,\ldots N$, in units in which $8\pi G \equiv M_{Pl}^{-2} = 1$.
Here $\bF^a = K^{a \bb} \bF_{\bb} = K^{a \bb} \bar{D}_{\bb} \bW $, where $D_a$ is the Kähler covariant derivative, $D_a W = \partial_a W + K_a W$, and
$K_{a \bb}$ is the Kähler metric.
We will use the shorthand $F_{a} \bar{F}^{a} \equiv F^2$.
%%v2 1/10  , so that $F$ is the magnitude of $F_{a} \bar{F}^{a}$.

We consider a set of critical points  $\{q\}$, satisfying
\bea
\partial_a V \large|_q= e^{K} \Bigl( {\cal D}_a(F_b) \bF^{b} - 2 F_a \bW  \Bigr) = 0 \, . \label{Vcrit}
\eea
Here and henceforth ${\cal D}_a$ denotes the appropriate Kähler and geometrically covariant derivative.
At any given point $q$ the scalar potential can be simplified by specifying the  Kähler gauge such that $\langle K \rangle_q=0$ and performing  an appropriate coordinate transformation such that $K_{a \bb} |_q = \delta_{a \bb}$.

\subsection{Matrix form of the critical point equation}
\label{sec:cpM}

The critical point equation \eqref{Vcrit} can be written as an eigenvalue equation of a particular Hermitian matrix, ${\cal M}$, formed from the second covariant derivatives of the superpotential
\cite{DD}.  Defining
\begin{equation}
Z_{ab} \equiv {\cal D}_a F_b\,,
\end{equation} and with $\vartheta_W \equiv {\rm Arg}(W)$, equation \eqref{Vcrit} can be expressed as
\be
{\cal M} \hF = 2 |W| \hF  \, , \label{eq:cp}
\ee
where
\be
{\cal M} =
\left(
\begin{array}{c c}
0 & e^{-i \vartheta_W} Z_{ab}  \\
e^{i \vartheta_W} \bZ_{\bar a \bb} & 0
\end{array}
\right) \, , \label{eq:M}
\ee
and the $2N$-dimensional vector $\hat F$ is given by
\be
\hF =
 \left(
\begin{array}{c}
e^{-i \vartheta_W} F^{\bb} \\
e^{i \vartheta_W} \bF^{b}
\end{array}
\right) \, .
\ee
The eigenvalues of ${\cal M}$ come in real pairs with opposite signs, $\pm \lambda_a$, with  $\lambda_a \geq 0$ and $a= 1, \ldots ,N$.  Thus, at any critical point, ${\cal M}$ must have an eigenvalue equal to $2 |W|$, and the vector $\hat{F}$ must be  proportional to the corresponding eigenvector. In \S\ref{ss:Wig} we will discuss the spectrum of eigenvalues of ${\cal M}$.

\subsection{Structure of the Hessian matrix}
\label{sec:Hess}
For a critical point $q$ to be a metastable vacuum, the eigenvalues of the Hessian matrix of the scalar potential evaluated at $q$ must all be positive. Denoting $U_{abc} = {\cal D}_a {\cal D}_b F_c\, $,
and writing the curvature of the field space\footnote{Our sign convention differs from that of \cite{DD}.} in terms of the partial derivatives of the Kähler potential as
$R_{a \bb c \bd} = \delta_{a \bar f }\ \partial_{c} \bar \Gamma^{\bar f}_{\bb \bd} = K_{a \bb c \bd} - K_{a c}^{~~ e} K_{\bb \bd  e}$, the
bosonic
mass matrix can be written
\bea
\partial^2_{a b} V &=& U_{a b c} \bF^c - \bW Z_{a b}  \,  , \\
\partial^2_{a \bb} V &=& \delta_{a \bb} \Big(F^2   - 2 |W|^2 \Big) - F_a \bar{F}_{\bb}- R_{a \bb c \bar d} \bar{F}^c F^{\bar d} + Z_{a}^{~\bc}\ \bar Z_{\bb \bar c} \, , \label{Vpp}
\eea where indices are raised and lowered with $\delta^{a\bar b}$.  The Hessian matrix ${\cal H}$ is thus given by
\bea
{\cal H} &=&
\left(
\begin{array}{c c}
\partial^2_{a \bb} V & \partial^2_{a b} V \\
\partial^2_{\ba \bb} V & \partial^2_{\ba b} V
\end{array}
\right)
\\
&=&
\left(
\begin{array}{c c}
 Z_{a}^{~\bar c}\ \bZ_{\bb \bc} -  F_a \bar{F}_{\bb} - R_{a \bb c \bar d} \bar{F}^c F^{\bar d}  & U_{a b c} \bar{F}^c - Z_{a b} \bW \\
\bU_{\ba \bb \bar c} F^{\bar c} - \bZ_{\ba \bb} W & \bZ_{\ba}^{~c}\ Z_{b c} -  F_b \bar{F}_{\ba} - R_{b \ba c \bar d} \bar{F}^c F^{\bar d}
\end{array}
\right) + \nonumber \\
&+&
\mathbb{1}  \Big(F^2 - 2 |W|^2 \Big) \,  ,
\eea
where $\mathbb{1}$ denotes the $2N \times 2N$  unit matrix.
The Hessian matrix is most conveniently analyzed in a `Goldstino' basis in which $F_{a} =  \delta_{a}^{~1} F e^{i \vartheta_F}  $.
In this basis the critical point equation \eqref{eq:cp} can be written
\be
Z_{11} = 2 |W|\ e^{i( 2 \vartheta_F - \vartheta_W)} \, , \qquad Z_{1 a'} = 0 \, ,
\ee
while the components $Z_{a' b'}$ remain unconstrained for $a', b' = 2, \ldots, N$. The Hessian matrix can be decomposed  into  constituent matrices as follows:
\be
{\cal H} = {\cal H}_{\rm susy} + {\cal H}_{\rm pure} + {\cal H}_{ K^{(4)}} + {\cal H}_{ K^{(3)}} + {\cal H}_{\rm shift} \, , \label{MasterHessian}
\ee
where
\bea
{\cal H}_{\rm susy}
&=&
\left(
\begin{array}{c c}
 Z_{a}^{~\bar c}\ \bZ_{\bb \bc} & 0 \\
0 &  \bZ_{\ba}^{~c}\ Z_{b c}   \end{array}  \label{firstconstituent}
\right) \, , \\
{\cal H}_{\rm pure}
&=&
\left(
\begin{array}{c c}
0 &  U_{a b 1} \bar{F}^1 - Z_{a b} \bW  \\
\bU_{\ba \bb \bar 1} F^{\bar 1} - \bZ_{\ba \bb} W & 0  \end{array}  \label{2constituent}
\right) \, , \\
{\cal H}_{ K^{(4)}}
&=&
F^2\ \left(
\begin{array}{c c}
- K_{a \bb 1 \bar 1} & 0\\
0 & - K_{b \ba 1 \bar 1}   \end{array}
\right) \, , \\
{\cal H}_{K^{(3)}}
&=&
F^2\  \left(
\begin{array}{c c}
K_{a 1}^{~~ e}\ K_{\bb \bar 1  e}  &  0  \\
0&K_{\bar{a} \bar{1}}^{~~\bar e}\ K_{b 1  \bar e}   \end{array}
\right)
\, , \\
{\cal H}_{\rm shift} &=& \mathbb{1} \Big(F^2-2|W|^2\Big)- F^2 \delta_{a}^{~1} \delta_{\bar
b}^{~\bar 1}- F^2 \delta_{\bar a}^{~\bar 1} \delta_{b}^{~1} \, .
\label{lastconstituent}
\eea
A few remarks are appropriate at this point.  First, ${\cal H}_{\rm susy}$ and ${\cal H}_{ K^{(3)}}$ are positive semidefinite.
Second, the mass scale $m_{\rm susy}$ of the supersymmetric masses is set by the
eigenvalues of ${\cal H}_{\rm susy}$, and can be larger or smaller than the scale $F$ that determines the soft supersymmetry-breaking masses.
The ratio $F/m_{\rm susy}$ (recall that we have set $\M=1$) has a significant effect on stability.  In \S\ref{sec:MMW} we will study generic critical points,
at which\footnote{See Appendix \ref{whatisgeneric} for a discussion of the distribution of $F/m_{\rm susy}$ in the set of all critical points.} $F \sim m_{\rm susy}$, and in \S\ref{sec:DD} we will consider `approximately-supersymmetric' critical points at which $F \ll m_{\rm susy}$.

In \S\ref{sec:RMT}, we will explain how the constituent matrices of $\cal{H}$ given in equations \eqref{firstconstituent}-\eqref{lastconstituent} can be identified as --- or well-approximated by --- elements of classical random matrix ensembles
with well-known emergent eigenvalue spectra at large $N$. The distribution of the eigenvalues of the Hessian matrix can then be obtained as the free convolution \cite{free} of the eigenvalue distributions of the constituent matrices, just as the distribution of a scalar random variable that is the sum of terms with known distributions can be obtained by the ordinary convolution of the constituent probability density functions.

\subsection{Defining a random supergravity}
\label{sec:RS}

To proceed further, we must specify the statistical properties of the entries of the matrices \eqref{firstconstituent}-\eqref{lastconstituent} constituting the Hessian matrix $\cal{H}$.  Our fundamental assumption --- consistent with that of \cite{DD} --- is that the components of tensors formed by covariant differentiation of $W$ and $K$  are {\it{independent, identically distributed (i.i.d.)\ variables}} drawn from some statistical distribution $\Omega$. We will occasionally abbreviate this by saying that $W$ and $K$ are
{\it{random functions}}.  Note that this assumption is quite different from taking the entries of $\cal{H}$ itself to be i.i.d.\ variables drawn from a distribution $\Omega$, which omits the structure and correlations implicit in (\ref{MasterHessian}).

We will use $\Omega(\mu,\sigma)$ to denote a complex\footnote{The diagonal elements of Hermitian matrices will of course be real.} distribution whose magnitude has mean $\mu$ and standard deviation $\sigma$, with a uniform distribution for the phase.  In \S\ref{universalityintro} we explain that as a consequence of the well-known phenomenon of universality in random matrix theory, the precise choice of $\Omega$ is immaterial, provided that the higher moments of $\Omega$ are appropriately bounded.

\subsubsection{The Kähler potential and its derivatives}

Suppose we took the Kähler potential to be a random function such that in a generic coordinate basis,
\begin{equation}
K_{a \bb}|_{q} \in \Omega(0,1) \,.
\end{equation}
Performing an orthogonal rotation to diagonalize $K_{a \bb}|_{q}$, the resulting eigenvalues will generically be of order $N$ (see \S\ref{sec:RMT} for details), and the $GL(N,\mathbb{C})$
transformation required to achieve $K_{a \bb} = \delta_{a \bb}$ involves rescaling by factors of order $N$.
To avoid performing this rescaling in all terms involving $K$, we find it convenient to take
\begin{equation}
K_{a \bb}|_{q} \in \Omega(0,{\scriptstyle \frac{1}{\sqrt{ N}}}) \,.
\end{equation}
Then, the $GL(N,\mathbb{C})$ transformation leading to $K_{a \bb} = \delta_{a \bb}$ does not involve any $N$-dependent rescalings.  More generally, the choice $\Omega(0,{\scriptstyle \frac{1}{ \sqrt{ N}}})$ is convenient because the random matrix eigenvalue spectra presented in \S\ref{sec:RMT} then have support in the same domain for all $N$.

Next, we need to specify the properties of  $K_{a \bb c}$ and $K_{a \bb c \bd}$ at $q$, in the basis in which $K_{a \bb} = \delta_{a \bb}$.
Ideally, the statistics of these objects would follow from a theory of general Kähler metrics (see e.g.\ \cite{Zelditch} for a very recent discussion of related issues), but for our purposes it will suffice to stipulate that the Kähler potential $K$ is a random function of its arguments, in the sense described above.
Then, $K_{a \bb c}$ and $K_{a \bb c \bd}$ do not take a special form in the basis in which the metric is diagonalized, and imposing the critical point equation \eqref{eq:cp} does not change this situation.
In the `Goldstino' basis in which $F_{a} = \delta_{a}^{~1} F\ e^{i \vartheta_F}$  at $q$, we have
\bea
K_{a \bb}|_{q}&=& \delta_{a \bb} \, ,  \\ \label{k1}
  K_{a \bb 1}|_{q} &\in& \Omega(0, {\scriptstyle \frac{1}{\sqrt{N}}}) \, , \\
  K_{a \bb 1 \bar 1}|_{q} &\in& \Omega(0, { \scriptstyle \frac{1}{\sqrt{N}}}) \, . \label{k3}
\eea

The assumptions (\ref{k1}),(\ref{k3}) are well-motivated for general Kähler manifolds, but we note that there are interesting exceptions, including the special geometry of the vector multiplet moduli space in ${\cal N}=2$ supergravity, for which the curvature tensor is given by
\be R_{a \bb c \bar d} =K_{a \bb c \bd} - K_{a c}^{~~e} K_{\bb \bd  e} =  K_{a \bb} K_{c \bar d} + K_{a \bar d} K_{c \bb} - e^{2K} K^{p \bar q}{\cal{F}}_{acp} {\cal{\bar{F}}}_{\bar b \bar d \bar q}\,,  \label{sg}
\ee
where ${\cal F}$ is the prepotential.  Repeating the analyses of \S\S\ref{sec:MMW},\ref{sec:DD} with the special geometry relationship (\ref{sg}) is straightforward, and we find a
%%v2
{\it{decreased}} likelihood of stability compared to the more general assumptions (\ref{k1}),(\ref{k3}) that are used throughout this work.

\subsubsection{The superpotential and its derivatives}

Turning now to the superpotential and its derivatives, we begin with a warmup in global supersymmetry.
Fixing a point $q$ in field space and working with canonically-normalized fields $\phi_A$, $A= 1,\ldots N$, we may write $W$ in the form (momentarily restoring factors of the Planck mass for clarity)
\begin{equation}
W = M^3 \,w(\phi_1/\M, \ldots ,\phi_N/\M) \,,
\end{equation} where $M$ is a mass scale.
Our assumption is that $w$ is a random function of its dimensionless arguments $x_A \equiv \phi_A/\M$, so that
\begin{equation}
\frac{\partial w}{\partial x_A}\Big|_{q} \in \Omega(\mu,\sigma) \,, \qquad  \frac{\partial^2 w}{\partial x_A \partial x_B}\Big|_{q} \in \Omega(\mu,\sigma)\,,
\end{equation} etc.
At a {\it{typical}} point the various derivatives of the superpotential will
%%v2
be of the same order of magnitude,
set by the physical effect responsible for the superpotential. (For example, in type IIB flux compactifications, the sizes of the superpotential and its derivatives are set by the flux scale.)  However, {\it{atypical}} points will play an important role in \S\ref{sec:DD}: it can happen that the superpotential, as well as its first derivative $F_a$,
are small compared to higher derivatives.  The result is a significant change in the stability criteria \cite{DD}.

In supergravity, the relevant expansion around $q$ is in (K\"ahler and geometrically) covariant derivatives of the superpotential.
We take
\bea
F_{a} \equiv {\cal D}_a W &\in& m_{\rm susy}  \  \Omega(0, { \scriptstyle \frac{1}{\sqrt{N} } })\, , \\
 Z_{a b} \equiv {\cal D}_a {\cal D}_b W &\in& m_{\rm susy} \ \Omega(0, { \scriptstyle \frac{1}{\sqrt{N}}}) \, , \\
 U_{a b c}  \equiv {\cal D}_a {\cal D}_b {\cal D}_c W &\in& m_{\rm susy} \ \Omega(0, {\scriptstyle \frac{1}{\sqrt{N}}}) \, .
\eea

We will assume that supersymmetry is spontaneously broken by an F-term; for a discussion of the possible effects of D-term energy (cf.~\cite{DD}), see \S\ref{sec:KKLTactual}.
The requirement of nonnegative vacuum energy then becomes
\begin{equation} \label{fw}
F \ge \sqrt{3}|W| \,,
\end{equation} so that in particular, $|W| \lesssim {\cal{O}}(F)$.
It is useful to define
\begin{equation} \label{defineomega}
\omega^2 \equiv \frac{3|W|^2}{F^2}\,,
\end{equation} so that $V=F^2(1-\omega^2)$, and equation (\ref{fw}) translates to $\omega \le 1$.
The stability properties of critical points depend on $\omega$, so we will repeat our analysis for a collection of fixed values of $\omega \in [0,1]$.  Thus, $|W|$ is determined in terms of $F$ and $\omega$.  For more details on this point, see Appendix \ref{whatisgeneric}.

We can now construct a model of a random supergravity by drawing $F_a$, $Z_{ab}$, $U_{abc}$, $K^{(3)}_{a \bb c}$, and $ K^{(4)}_{a \bb c \bd}$ independently from the distributions specified above, at each point $q$.
As we are primarily interested in critical points, we should study the set of points $\{q\}$ subject to the critical point equation \eqref{eq:cp}.  Such points are {\it{not}} completely generic: equation \eqref{eq:cp} enforces a particular correlation between $Z_{ab}$, $W$, and $F_a$, as reviewed in the discussion following equation \eqref{eq:cp}.  Following \cite{DD}, we will carefully incorporate the restriction implied by equation \eqref{eq:cp}.

\section{Random Matrix Theory for Supergravity}
\label{sec:RMT}

In this section we will briefly review a few important concepts and results from random matrix theory, in order to make our analysis more self-contained.
An accessible and fairly recent introduction can be found in \cite{Rao}; see also the text by Mehta \cite{Mehta}.

% %%%%%%%%%%%%%%%%%%%%%%%%%%%%%%
%%%%%%%%%%%%%%%%%%%%%%%%

\subsection{Classical ensembles} \label{sec:RMTensembles}

A foundational idea in random matrix theory is that given only limited statistical information about the entries of a diagonalizable $N \times N$ matrix, for large $N$ one can make incisive statements about the statistical properties of the eigenvalues.  For our purposes, the properties of principal interest are the eigenvalue spectrum for a typical matrix, and the probability of a large fluctuation of the smallest eigenvalue.

We begin by reviewing the ensembles relevant for this work.
%Although the ensembles presented here have been extensively studied for real, complex, and quaternionic entries,  we present results only for the complex case of direct interest.

\subsubsection{The Wigner ensemble} \label{ss:Wig}

One of the simplest and best-known ensembles of random matrices is the {\it{Wigner ensemble}} of
Hermitian matrices, also referred to as the Gaussian Unitary Ensemble \cite{Wigner0,Wigner,Wigner1}.  Elements of this ensemble, which we refer to as Wigner matrices, are $N\times N$ Hermitian matrices $M$ given by
\begin{equation}
M= A+A^{\dagger}  \, ,
\end{equation} where $A_{ij}$ for $i,j = 1,\ldots, N$ are i.i.d.\ variables drawn from
$\Omega(0,\sigma)$, and the dagger denotes Hermitian conjugation.

The measure on the space of matrices is
\be
dP(M) = \prod_{1\leq i \leq j \leq N} \ f(M_{ij})\ d M_{ij} \, ,
\ee
where $f(M_{ij})$ denotes the probability density of observing $M_{ij}$. For normally-distributed entries of  $M$, the joint probability density of the eigenvalues $\lambda_1, \ldots, \lambda_N$ is obtained by a unitary change of coordinates,
\begin{equation}
f(\lambda_1,\ldots,\lambda_N) = {\cal{C}}\ {\rm{exp}}\Bigl(-\frac{1}{\sigma^2}\sum_{i=1}^{N} \lambda_{i}^2 +  2 \sum^N_{i < j}{\rm{ln}}|\lambda_i-\lambda_j| \Bigr)\,,
\end{equation}
where  ${\cal{C}}$ is an $N$-dependent normalization constant. As  conceived in the famous work of Dyson \cite{Dyson1}, this joint probability density can be given a physical interpretation in terms of a one-dimensional Coulomb gas of $N$ charged particles executing Brownian motion under the influences of a  confining quadratic potential and of mutual electrostatic repulsion. This physical picture has proved to be very fruitful in deriving exact results for a variety of properties of the eigenvalue spectrum (see e.g.~\cite{Dean2,Dean}), and in \S\ref{sec:MMW} and \S\ref{sec:DD} we will see that repulsion between pairs of eigenvalues significantly impacts the stability of critical points in  supergravity.

\begin{figure}
\begin{center}
$\begin{array}{l c r}
\includegraphics[width=8.2cm]{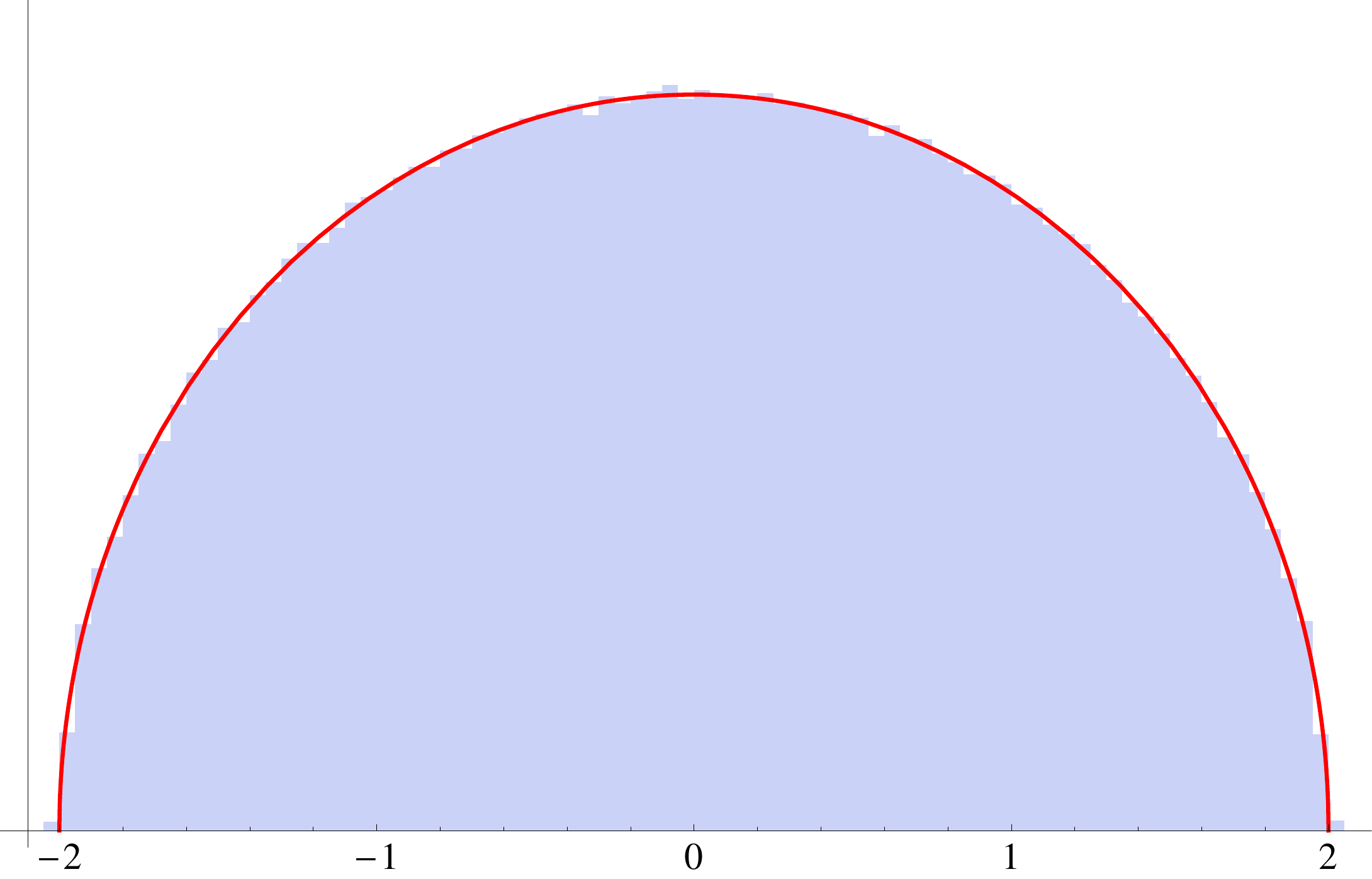} & \includegraphics[width=8.2cm]{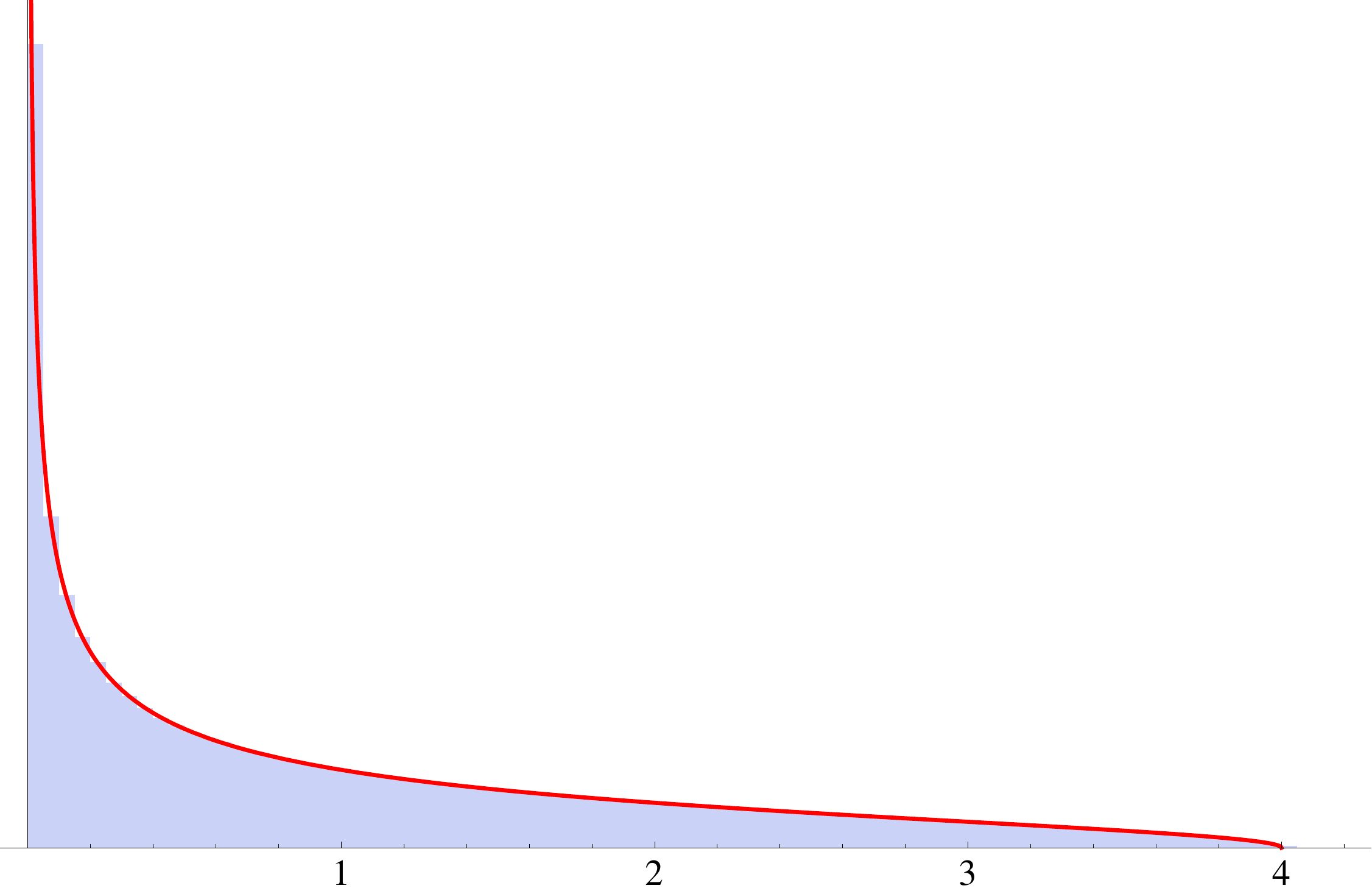}
\end{array}$
\caption{The eigenvalue spectra for the Wigner ensemble (left panel), and the Wishart ensemble with $N=Q$ (right panel), from $10^3$ trials with $N=200$.}
\label{fig_WigWish}
\end{center}
\end{figure}

At large $N$,
the eigenvalue spectrum
of a Wigner matrix converges to the celebrated Wigner semicircle law,
\begin{equation} \label{wigs}
\rho(\lambda) =  \frac{1}{2\pi N \sigma^2}\sqrt{4N \sigma^2-\lambda^2} \,.
\end{equation} where $\rho$ is the eigenvalue density.  Setting $\sigma=\frac{1}{\sqrt{N}}$, the eigenvalue spectrum has support in the interval $[-2,2]$, cf.\ Figure \ref{fig_WigWish}.

\subsubsection{The Wishart ensemble}

The second class of random matrices we will need are {\it{complex Wishart matrices}}, which take the form
\begin{equation}
M=A A^{\dagger}\,,
\end{equation}
where $A$ is an $N\times Q$ complex matrix with entries drawn from $\Omega(0,\sigma)$, and $Q \ge N$.
The study of this ensemble dates back to Wishart's investigation of sample covariance matrices \cite{Wishartpaper}, and the universality evident in the Wishart ensemble provided some of the inspiration for Wigner's subsequent development of random matrix theory.

As a Wishart matrix is the Hermitian square of another matrix, it is necessarily positive semidefinite. The joint probability density of a complex Wishart matrix is (cf.~e.g.\ \cite{Rao})
\be
f(\lambda_1, \ldots, \lambda_N) = {\cal C}\ \exp\Big(- \frac{1}{\sigma} \sum_{i=1}^N \lambda_i \ + 2\sum_{i < j}^N{\rm{ln}}|\lambda_i - \lambda_j| + (Q-N)\,\sum_{i}^N {\rm{ln}} \lambda_i \Big) \, .
\ee
In the Coulomb gas picture, the non-negativity of a Wishart matrix corresponds to the presence of a hard wall at $\lambda=0$.

The eigenvalue distribution in the Wishart ensemble
is given by the  Mar\v{c}enko-Pastur law \cite{MP}, which takes the form
\begin{equation} \label{wishs}
\rho(\lambda) =  \frac{1}{2\pi N \sigma^2 \lambda}\sqrt{(4N\sigma^2-\lambda)\lambda} \,,
\end{equation} for the special case $N=Q$ that will be relevant in our analysis, cf.\ Figure \ref{fig_WigWish}.

The probability density function of the smallest eigenvalue $\lambda_1$ was first computed by Edelman \cite{Edelman}, and for our purposes it suffices to note that for $N=Q$ and $\sigma = \frac{1}{\sqrt{N}}$, its average position $\langle\lambda_{1}\rangle$
scales
as $\frac{1}{N^2}$.

\subsubsection{The Altland-Zirnbauer C$I$ ensemble}
\label{sec:CI}
\begin{figure}
\begin{center}
$\begin{array}{l c r}
\includegraphics[width=8.2cm]{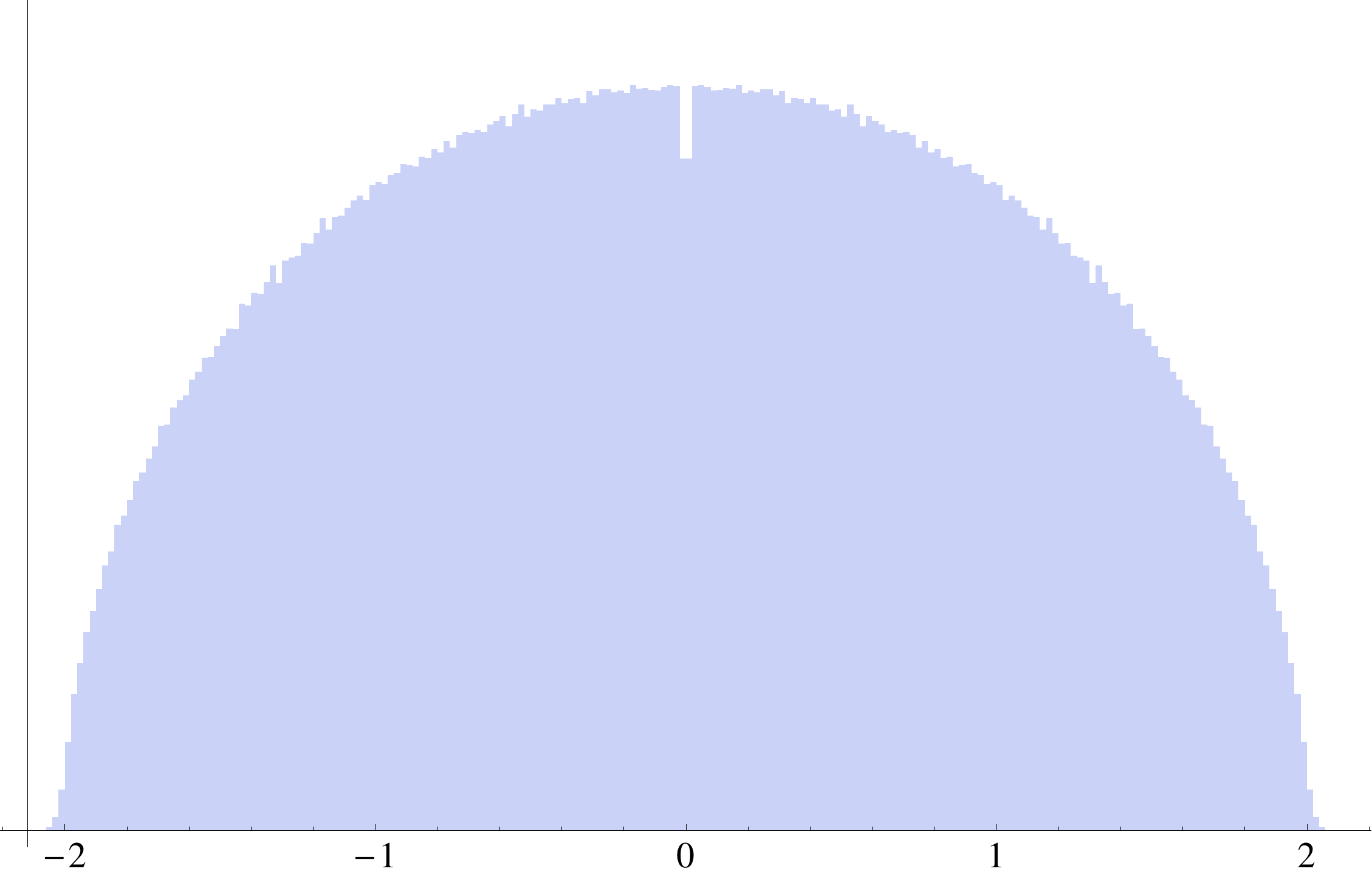} & \includegraphics[width=8.2cm]{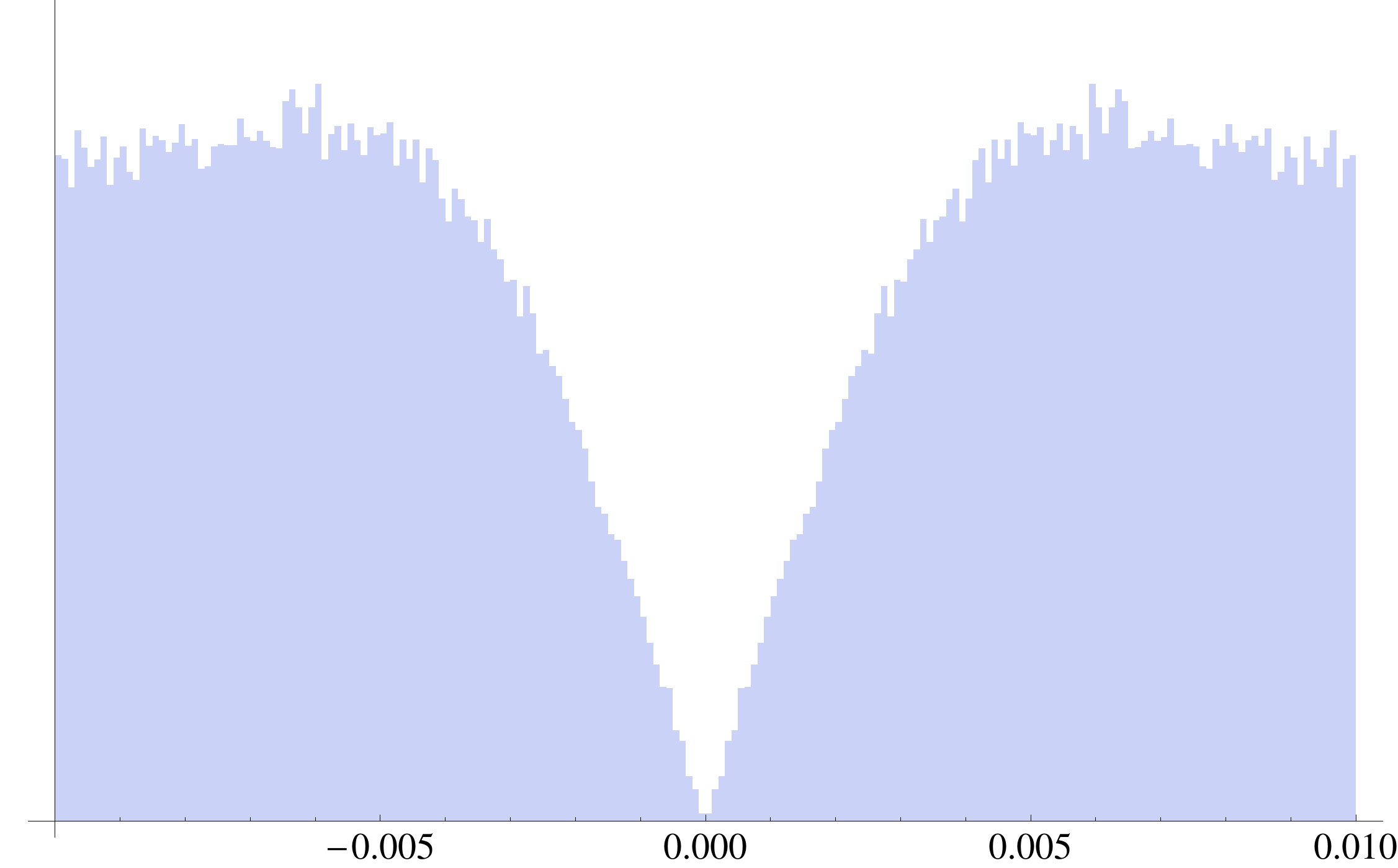}
\end{array}$
\caption{The eigenvalue spectrum for the C$I$ ensemble,  from $10^5$ trials with $N=200$.  The full spectrum appears in the left panel, while the right panel shows the details of the cleft at $\lambda=0$.  Notice that the boundary of the linear regime occurs for $\lambda \sim \frac{1}{N}$.}
\label{fig_CI}
\end{center}
\end{figure}

The matrix ${\cal M}$ appearing in the critical point equation (\ref{eq:cp}) has an eigenvalue spectrum that is broadly reminiscent of the Wigner semicircle law, but the $2N$ eigenvalues of ${\cal M}$ come in opposite-sign pairs $\pm \lambda_a$, with  $0\le \lambda_1 \le \ldots \le \lambda_{N}$.
As observed in \cite{DD}, matrices ${\cal M}$ of the form (\ref{eq:M}) belong to the {\it{Altland-Zirnbauer C$I$ ensemble}} \cite{AZ}.
For normally-distributed entries of ${\cal M}$, the joint probability density of the eigenvalues  is
\be
f(\lambda_1,\ldots,\lambda_N) = {\cal{C}}\ {\rm{exp}}\Bigl(-\frac{1}{\sigma^2} \sum_{i=1}^{N} \lambda_{i}^2+  \sum^N_{i\neq j}{\rm{ln}}|\lambda^2_i-\lambda^2_j| + \sum^N_{i=1} \ln |\lambda_i|\Bigr)\,. \label{eq:CI}
\ee
In the Coulomb gas picture,  the additional term $\sum^N_{i=1} |\lambda_i|$ can be interpreted as encoding a repulsive force between each mirror pair of eigenvalues,
$\pm\lambda_i$ \cite{AZ}. This repulsion is particularly important for the smallest eigenvalue of ${\cal M}$, and leads to a linear cleft in the eigenvalue spectrum for small $\lambda$:
\begin{equation}
\rho(\lambda) \approx  k \lambda + {\cal{O}}(\lambda^3) \,,
\label{eq:Cleft}
\end{equation} with $k$ a constant of order unity,
so that the eigenvalue density vanishes at $\lambda=0$,  cf.\ Figure \ref{fig_CI}.
Recalling that the critical point equation \eqref{eq:cp} requires that ${\cal M}$ has $2|W|$ as an eigenvalue, we see that critical points with very small $|W|$ are rare in comparison to those with $2 |W| \approx 1$.

\subsection{Fluctuations of extreme eigenvalues} \label{extremereview}

The eigenvalue spectra presented above describe the {\it{typical}} configurations of eigenvalues: for example, the eigenvalue spectrum for an ensemble of Wigner matrices with entries drawn from $\Omega(0,\sigma)$ is zero outside $[-2\sqrt{N}\sigma,2\sqrt{N}\sigma]$,  but this does not mean that no matrix in the ensemble has eigenvalues outside this range.  Instead, the Wigner spectrum has a `soft edge'  at each end of the semicircle: there is a nonzero probability that one or more eigenvalues can be found beyond this edge.  In contrast, the Wishart spectrum has a `hard edge' at $\lambda=0$, as the matrices in question are necessarily positive semidefinite, while the other edge of the spectrum is soft.

As we shall soon establish, the spectrum of the Hessian matrix ${\cal H}$ extends to negative values, so that the presence of tachyons is generic, but not guaranteed.  It is therefore essential to determine the probability that the smallest eigenvalue of  ${\cal H}$ happens to be large enough so that ${\cal H}$ is positive definite.  We refer to this occurrence as a fluctuation to positivity.

\subsubsection{Eigenvalue fluctuations and the Tracy-Widom law}

The study of fluctuations of the smallest (or largest) eigenvalue was initiated in the pioneering work of Tracy and Widom \cite{TW}; see also \cite{Forrester}.
The theory of fluctuations is best-developed when the fluctuations are suitably small, with the deviation of the extreme eigenvalue from its mean position being ${\cal O}(N^{-1/6})$ for the case of the Wigner ensemble.  In this case,
the smallest eigenvalue $\lambda_{1}$ is given by (cf.\ \cite{Nadal} for a useful summary)
\begin{equation}  \label{tweq}
\lambda_{1} \approx -2\sqrt{N} + N^{-1/6}\chi
\,,
\end{equation}
where for $N \to \infty $, $\chi$ is a random variable that  follows the Tracy-Widom distribution $F_2$ \cite{TW}.
Extension of the Tracy-Widom law to the Wishart ensemble was achieved for real matrices in \cite{Johnstone}, and for complex Wishart matrices in \cite{Johansson}.

In \S\ref{sec:MMW} we will find that for a typical supergravity critical point, a fluctuation of size ${\cal O}(N^{-1/6})$ of the smallest eigenvalue of ${\cal H}$ is insufficient to render ${\cal H}$ positive definite.  We therefore require an extension of the Tracy-Widom theory describing large fluctuations, with the deviation from the mean position being as large as ${\cal O}(\sqrt{N})$.
The theory of large fluctuations has been developed\footnote{Earlier work on related fluctuations appears in \cite{glass}.  For applications to counting critical points of random functions, see e.g.\ \cite{Bray}.} in a series of works by Majumdar and collaborators \cite{Dean2,Vivo,Dean,Vergassola,Nadal2,Nadal},
which we now briefly review, focusing on the Wigner ensemble.

Through a saddle point computation of the partition function in the Coulomb gas model, Dean and Majumdar \cite{Dean2,Dean} were able to evaluate the probability of a large fluctuation of the smallest eigenvalue $\lambda_1$ of a Wigner matrix to the right of its mean position $\langle \lambda_{1} \rangle \equiv-2\sqrt{N}$.  The result, at leading order in large $N$, is \cite{Dean2,Dean}
\begin{equation}  \label{dmeq}
P\Bigl(\lambda_1\ge t\Bigr) \propto  {\rm{exp}}\Bigl[-2N^2 \psi_{-}\left(t/\sqrt{2 N}\right) \Bigr] \,,
\end{equation} for $t\ge -2\sqrt{N}$ and $t+2\sqrt{N} \sim {\cal{O}}(\sqrt{N})$.
Although $\psi_{-}(y)$ is known in closed form, we present here only the result relevant for a fluctuation to positivity:
\begin{equation}
\psi_{-}(0) = \frac{{\rm{ln}}(3)}{4} \,.
\end{equation}
(The corresponding result of \cite{Dean2,Dean} for {\it{real}} Wigner matrices
agrees very well with the earlier numerical results of Aazami and Easther \cite{Aazami:2005jf} for the same ensemble.)  In summary, the probability that an $N\times N$ complex Wigner matrix is positive-definite is given by \cite{Aazami:2005jf,Dean2,Dean}
\begin{equation}  \label{dmres}
P \propto {\rm{exp}}\Bigl[-c N^2 \Bigr] \,,
\end{equation}
with $c\approx \frac{{\rm{ln}}(3)}{2}$.
At large $N$ this is exceptionally small compared to the estimate $P \approx 2^{-N}$ that follows from the naive assumption that the $N$ eigenvalues are independent.  Of course, eigenvalue interactions are fundamental to random matrix theory, so it is no surprise that omitting these interactions gives an entirely inaccurate result for the probability of positivity.

An intuition from the Coulomb gas model will be helpful in our analysis.  Consider a fluctuation of the smallest eigenvalue $\lambda_{1}$ to roughly the midpoint of the spectrum, as would be required for a fluctuation to positivity in the Wigner ensemble.  The distance involved is ${\cal{O}}(\sqrt{N})$, and ${\cal{O}}(N)$ eigenvalues need to be displaced.  As these eigenvalues experience a quadratic potential, the total energetic cost
is ${\cal{O}}(N^2)$, consistent with the detailed results of \cite{Dean2,Dean}.
Similar results have been obtained for inward fluctuations of the soft and hard edges of the Wishart spectrum in
\cite{Vivo} and \cite{KC}, respectively.  The lesson is that a substantial inward shift of one or more eigenvalues has a statistical cost $\sim {\rm{exp}}(-N^2)$, and is hence extremely unlikely at large $N$.

Having assembled the necessary tools, we now turn to studying  the stability of the Hessian.

%%%%%%%%%%%%%%%%%%%%%%%%%%%%%%%%%%%%%%%%%%%%%%%%%%%%%%%
%4
%%%%%%%%%%%%%%%%%%%%%%%%%%%%%%%%%%%%%%%%%%%%%%%%%%%%%%%
\section{Stability of Generic Critical Points}
\label{sec:MMW}

In this section we will study the Hessian ${\cal H}$ at a generic critical point, where
\be \label{genericpoint}
 F_a \sim  Z_{a b} \sim U_{a b c} \, .
 \ee
We refer the reader to Appendix \ref{whatisgeneric} for a detailed demonstration that such points are indeed generic.

In \S\ref{simplemodel} we examine the decomposition (\ref{MasterHessian}) of the Hessian matrix into a sum of terms and argue that each term is well-approximated as a member of one of the classical ensembles reviewed in \S\ref{sec:RMTensembles}.
We then obtain the eigenvalue spectrum analytically from the free convolution of the constituent spectra.
In \S\ref{MMWstability} we argue that the probability that a given critical point is a metastable vacuum can be obtained by adapting the results of \cite{Dean2,Dean} to the free convolution model.
We then perform an extensive numerical analysis of the full Hessian matrix, finding that a generic critical point is exponentially unlikely to be a metastable vacuum.  Thus, despite the abundance of critical points, this region of the random supergravity landscape is indeed a wasteland.

\subsection{The Hessian spectrum from a free convolution}
\label{simplemodel}

The Hessian ${\cal H}$ can be decomposed according to \eqref{MasterHessian} as ${\cal H} = {\cal H}_{\rm susy} + {\cal H}_{\rm pure} + {\cal H}_{ K^{(4)}} + {\cal H}_{ K^{(3)}} + {\cal H}_{\rm shift}$.

From \S\ref{sec:RMT} we immediately recognize that each of ${\cal H}_{\rm susy}$ and ${\cal H}_{K^{(3)}}$ is very similar to a double copy of an $N$-dimensional complex Wishart matrix.
The correspondence is imperfect because $Z$ (respectively $K^{(3)}$) is symmetric, so that the diagonal entries have twice the variance of the off-diagonal entries.
We have verified that this minor difference does not significantly affect the eigenvalue spectrum.

Next, ${\cal H}_{\rm pure}$ and ${\cal H}_{K^{(4)}}$ can be modeled as $2N$-dimensional Wigner matrices.  Once again, this is an approximation: the actual Hessian matrix at a critical point must incorporate the critical point constraint \eqref{eq:cp}.  The sum of two Wigner matrices is again a Wigner matrix, so we may write ${\cal H}_{\rm pure}+{\cal H}_{K^{(4)}} \approx \rm{Wigner}$.

Assembling the pieces, and noting
%% v2 that ${\cal H}_{\rm shift}$ simply translates the eigenvalue spectrum, our model amounts to
that the effect of ${\cal H}_{\rm shift}$ on the bulk of the eigenvalue spectrum is simply a translation, our model amounts to
\begin{equation}
{\cal H} \approx \rm{Wigner}({\cal H}_{\rm pure}+{\cal H}_{K^{(4)}}) + \rm{Wishart}({\cal H}_{\rm susy}) + \rm{Wishart}({\cal H}_{K^{(3)}})\,.
\end{equation}

\subsubsection{Free convolutions and sums of random matrices} \label{free}

To obtain the spectrum, we need to address a problem of the general form: ``if A and B are random matrices with known eigenvalue spectra $\mu_A$, $\mu_B$, what is the spectrum $\mu_{A+B}$ of their sum A+B?"  If A and B were to commute, $\mu_{A+B}$ would be the convolution of $\mu_A$ and $\mu_B$, but there is no justification for this assumption in our case.  The solution of the general problem is provided by Voiculescu's theory of free probability.
We will describe here only the immediately relevant tools of free probability, referring the reader to the text \cite{free} for details and references.

Given two ensembles A, B of random matrices, the {\it{free convolution}} $\boxplus$ is defined such that
\begin{equation} \label{definefree}
\mu_{A}\boxplus\mu_{B} = \mu_{A+B}\,,
\end{equation} i.e.\ the free convolution of the spectra of the summands is the spectrum of the sum.  Just as cumulants are additive under ordinary convolution of random variables, free cumulants can be defined with the same additivity property under the free convolution.
In principle $\mu_{A+B}$ can be obtained from the R-transform, which is the generating function of the free cumulants \cite{Speicher}, but inversion of the R-transform can be rather cumbersome.
For the large class of \emph{algebraic} random matrices \cite{RaoPoly}, which includes sums of Wigner and Wishart matrices, a more efficient approach \cite{MP,RaoPoly} is to work with the
Stieltjes transform \cite{RaoPoly}.

The Stieltjes transform of a probability measure ${\rm d}\mu(x) = \rho(x) dx$ with support on a real interval $I$ is defined by
\be
m_{\mu}(z) = \int_{I} \frac{{\rm d}\mu(x)}{z-x}\, ,
\end{equation}
where ${\rm Im}(z) > 0$. Algebraic random matrices are random matrices for which $m_{\mu}(z)$ is the solution of a polynomial equation in $m_{\mu}$ and $z$, e.g.~the Stieltjes transform of the Wigner density solves the equation
\be
m^2_{\mu} + a\ z m_{\mu} + a = 0 \, ,
\ee
where $a= \big( N \sigma^2\big)^{-1}$.
The probability density is readily obtained from $m_{\mu}(z)$ using the Stieltjes-Perron inversion formula,
\be
\rho(x) = \frac{1}{\pi}\ \lim_{\epsilon \rightarrow 0} \ {\rm Im} \ m_{\mu}(x + i \epsilon) \, .
\ee
Edelman and Rao have shown that the free convolution can be implemented efficiently through manipulations of polynomials involving the Stieltjes transform  \cite{RaoPoly}.

\subsubsection{The spectrum as  Wigner $\boxplus$ Wishart $\boxplus$ Wishart} \label{freeresult}
\begin{figure}
\begin{center}
$\begin{array}{l c r}
\includegraphics[width=10.2cm]{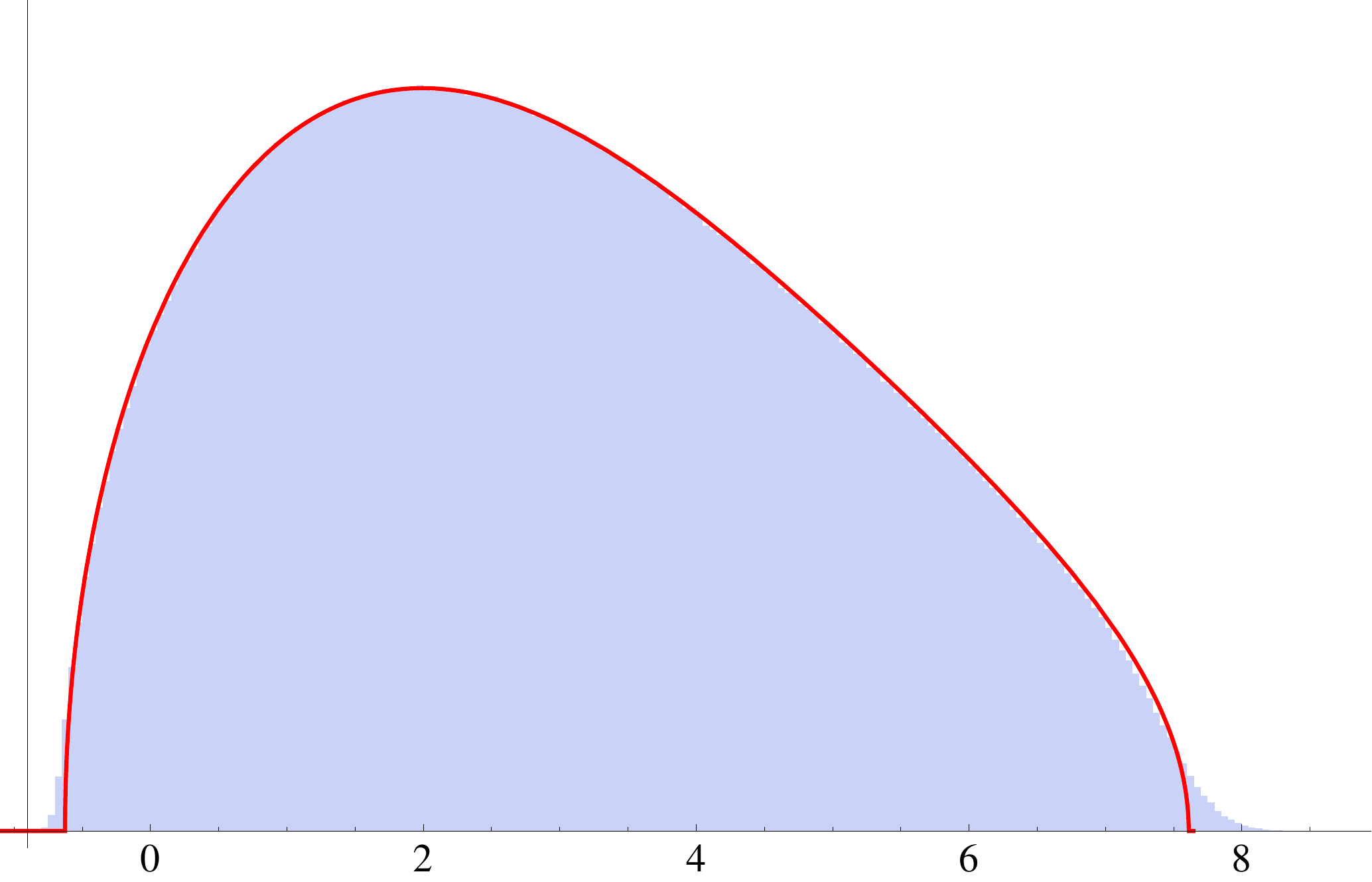}
\end{array}$
\caption{The histogram shows the spectrum of eigenvalues of the full Hessian matrix ${\cal H}$ (\ref{MasterHessian}) for $N=200$ and $\omega=.1$, in units of $F^2$, while the curve gives the analytic result (\ref{MasterSpectrum}) from the Wigner $\boxplus$ Wishart $\boxplus$ Wishart model, with no adjustable parameters.}
%%v2 1/10 caption includes units
%%v2 1/12 \omega = .1
\label{fig_WWW}
\end{center}
\end{figure}

In terms of the free convolution  $\boxplus$ defined in equation (\ref{definefree}), we may write the eigenvalue spectrum $\rho({\cal{H}})$ as
\begin{equation}
\rho({\cal{H}}) \approx \rho(\rm{Wigner}) \boxplus \rho(\rm{Wishart}) \boxplus \rho(\rm{Wishart})\,,
\end{equation} where $\rho(\rm{Wigner})$  is given in equation (\ref{wigs}) and $\rho(\rm{Wishart})$ is given in equation (\ref{wishs}).
%%v2
Obtaining the
Stieltjes transforms of $\rho(\rm{Wigner})$ and $\rho(\rm{Wishart})$ and using the polynomial method of \cite{RaoPoly}, we find the spectrum

\begin{equation} \label{MasterSpectrum}
\rho(\lambda) =  \frac{ 3 \omega^4 +117 +9 \left( \omega^2 - 4\right) \lambda+ 9 \lambda^2 -\left(\frac{3}{2}\right)^{2/3} \psi(\lambda)^{2/3}  }{2^{2/3}\ 3^{5/6}\ \pi\
   \psi(\lambda)^{1/3}\ \left(\omega^2+6\right)}\,,
\end{equation} where $\omega = \frac{\sqrt{3} |W|}{F}$, $\psi(\lambda)$ is given by
\bea
 \psi(\lambda) &=& 9 \omega^4 (\lambda+1)+27 \omega^2 \left(\lambda^2-5 \lambda+24\right)
-\sqrt{3 \tau}+18 \lambda^3-108 \lambda^2+216 \lambda+1314\,,
\eea
and
\bea
\tau &=&  -\left(\omega^2+6\right)^2 \Bigl[ 81 \lambda ^4+162 \lambda ^3 \left(\omega ^2-7\right)+9 \lambda ^2 \left(13 \omega
   ^4-180 \omega ^2+621\right)+\\ \nonumber
  &+& 18 \lambda  \left(2 \omega ^6-35 \omega ^4+333 \omega
   ^2-630\right)+4 \omega ^8-48 \omega ^6+873 \omega ^4-12636 \omega ^2-9396\Bigr]\,.
\eea
%%v2
This is one of our primary results.
Figure \ref{fig_WWW} illustrates the
%%v2 1/12 surprisingly good
remarkably good agreement between %our analytic result
(\ref{MasterSpectrum}) and simulations of the full ${\cal H}$.

\subsection{Eigenvalue fluctuations and de Sitter vacua} \label{MMWstability}

Although we now have an analytic result for the eigenvalue spectrum in the Wigner $\boxplus$ Wishart $\boxplus$ Wishart model, which gives an excellent approximation to the spectrum of ${\cal H}$ itself, computing the probability of a large fluctuation for this model is rather involved.  The literature summarized in \S\ref{extremereview} contains detailed characterizations of the fluctuations of extreme eigenvalues of Wigner or Wishart  matrices, but a direct computation of the large fluctuations in the Wigner $\boxplus$ Wishart $\boxplus$ Wishart model would require a dedicated saddle point analysis along the lines of \cite{Dean2,Dean}, and is beyond the scope of the present work.

From the Coulomb gas model, it is clear that a sufficiently large fluctuation of the smallest eigenvalue will be sensitive to the global shape of the spectrum: such a fluctuation will displace the eigenvalues to its right, with an energy cost that depends on their density.  However, a small fluctuation will displace only the eigenvalues near the edge of the spectrum, and the likelihood of such a fluctuation should therefore depend only on the edge shape.
Correspondingly, it has been conjectured \cite{RaoPoly} that Tracy-Widom fluctuations will be seen in essentially any algebraic random matrix whose eigenvalue density has square root behavior at its edge.  As our Wigner $\boxplus$ Wishart $\boxplus$ Wishart model falls in this class, we expect that {\it{small}} fluctuations of the smallest eigenvalue of  ${\cal H}$ will be governed by the Tracy-Widom law.

Examining the position of the left edge in the Wigner $\boxplus$ Wishart $\boxplus$ Wishart model, we see that a fluctuation to positivity is {\it not} a small fluctuation in the sense of \cite{TW}: the distance to the origin is\footnote{For ease of comparison to \cite{TW}, we take $\Omega = {\cal{N}}(0,1)$ in this discussion, although we set $\Omega = {\cal{N}}(0,{ \scriptstyle \frac{1}{\sqrt{N} } })$ elsewhere.} ${\cal O}(\sqrt{N})$, not ${\cal O}(N^{-1/6})$.  However, between the left edge and the origin, the spectrum has a shape very reminiscent of the semicircle law.  Emboldened by this, we anticipate that
the probability of a large fluctuation of the smallest eigenvalue of ${\cal H}$ is accurately modeled using the corresponding probability (\ref{dmres}) in the Wigner ensemble, i.e.\ we expect $P \propto {\rm exp}(-c\,N^p)$  with $p \sim 2$.

\subsection{Numerical results}

We now report on the results of extensive simulations of fluctuations to positivity in the full ${\cal H}$ model.  These simulations make no approximation.  We include the full structure of ${\cal H}$ (e.g., the slight difference between ${\cal H}_{\rm susy}$ and a Wishart matrix), and we do not rely on any of the analytical results reviewed in \S\ref{sec:RMT}.  No expansion in $\frac{1}{N}$ or in $F$ is used.  We simply create an ensemble of realizations of ${\cal H}$, following the prescription of \S\ref{sec:cp}, and directly determine the fraction that are positive definite.

Naturally, these simulations could still fail to yield an accurate picture of the positivity probability in the supergravities derived from string theory: in particular, our definition of a random supergravity could be non-representative.  Moreover, computational cost imposes an upper limit on $N$, and our extrapolation to larger $N$ could be inaccurate.

One detail of ${\cal H}$ requires further explanation.  The critical point condition, equation (\ref{eq:cp}), enforces that ${\cal M}$ has eigenvalue $2|W|$.  For a given $W$, no matrix drawn randomly will have a precisely correct eigenvalue (this reflects the fact that critical points are a measure zero subset of all points).  To impose this constraint, we note that if ${\cal M}$ has an eigenvalue $\lambda_{W} \in [(2-2\epsilon)|W|, (2+2\epsilon)|W|]$, but $\lambda_{W} \neq 2|W|$, the distortions of the spectrum compared to that found at a genuine critical point will be of order $\epsilon$.  In our simulations, we have taken $\epsilon=10^{-2}$.
%, so that some ${\cal M}$ are discarded as incompatible with equation (\ref{eq:cp}).  The
In \S\ref{sec:DD} this issue will pose a greater challenge: for $|W| \ll 1/N$, a very small fraction of randomly drawn ${\cal M}$ will fall in $[(2-2\epsilon)|W|, (2+2\epsilon)|W|]$, cf.~equation (\ref{eq:Cleft}),
%obey the constraint,
and it becomes computationally costly to find examples.

\begin{figure}
\begin{center}
$\begin{array}{l c r}
\includegraphics[width=13.5cm]{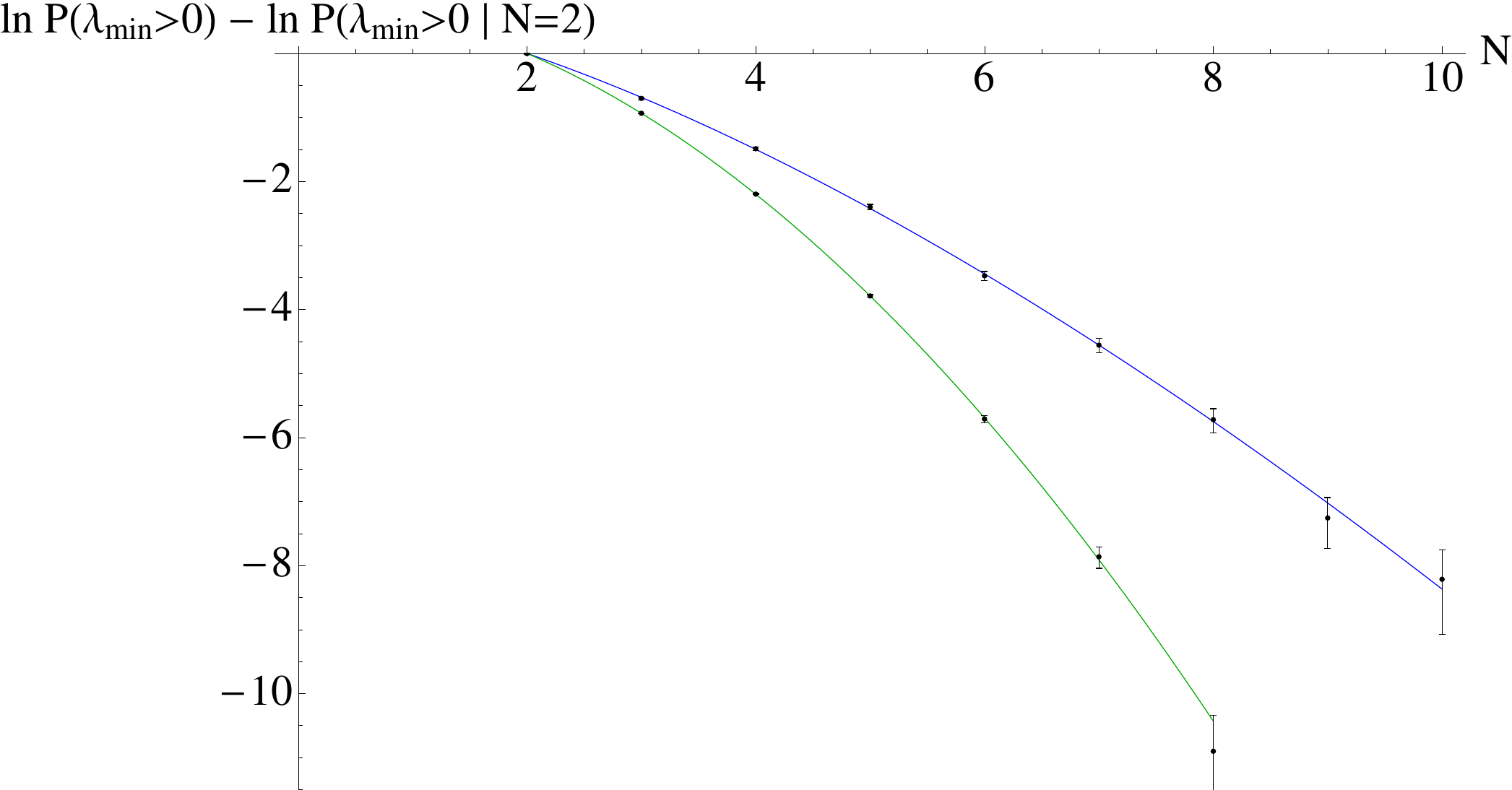}  %%v2 1/10 figure axis label changed
\end{array}$
\caption{The logarithm of the probability $P(\lambda_{\rm min}>0)$ that the smallest eigenvalue of ${\cal H}$ is positive, as a function of $N$, with $\omega=1$. Upper branch: simulations of the full Hessian matrix ${\cal H}$,  with best-fit values $p=1.50 \pm 0.10$, $c=0.29 \pm 0.06$.  Lower branch: simulations of the Wigner $\boxplus$ Wishart $\boxplus$ Wishart model,  with best-fit values $p=1.90 \pm 0.04$, $c=0.21 \pm 0.02$.
The error bars give the 2$\sigma$ statistical uncertainty.}
\label{fig_MMWWWW}
\end{center}
\end{figure}

%%v2 1/12
Figure \ref{fig_MMWWWW} presents the results of simulations of the full Hessian matrix ${\cal H}$ (upper curve) and of the  Wigner $\boxplus$ Wishart $\boxplus$ Wishart model (lower curve).
The qualitative properties are similar, but the best-fit values of $p$ are somewhat different.  This is not surprising, as the numerically-accessible values of $N$ are not large: for $N \gtrsim 20$, stability is extremely rare, and it is difficult to obtain sufficient statistics to characterize the probability of positivity.  For $N \lesssim 20$, the large $N$ expansion underpinning our random matrix theory approach is marginal at best, with two important consequences.  First, the correspondence between the model and simulation spectra is imperfect for $N \sim 20$ (contrast the superb agreement for $N=200$ shown in Figure \ref{fig_WWW}), and the corresponding difference between the left edges of these spectra contributes to a different fluctuation probability.  A primary cause of the difference between the spectra of the
%%v2 1/12 the
full Hessian matrix ${\cal H}$ and
%%v2 1/12
of the analytical model (\ref{MasterSpectrum}) is that ${\cal H}_{\rm pure}$ involves the matrix $Z$, as does ${\cal H}_{\rm susy}$, so that in the Wigner $\boxplus$ Wishart $\boxplus$ Wishart model, the Wigner matrix is correlated with one of the Wishart matrices.  For $N \gg 1$ (and also for $\omega \ll 1$) this correlation becomes less important.  Second, for small enough $N$, fluctuations to positivity are governed by the Tracy-Widom law (\ref{tweq}) rather than by the considerably steeper large-fluctuation expression (\ref{dmeq}), so that a fit that includes data points starting from $N=2$ will result in a value of $p$ that is {\it{smaller}} than the asymptotic large $N$ value.

In summary, due to the challenges inherent in studying extremely rare events numerically, we have not obtained sufficient data at large $N$ to make a definitive determination of the large $N$ behavior of the probability, and this is an interesting problem for the future.  In light of the arguments of \S\ref{MMWstability}, it remains reasonable to conjecture that $p \sim 2$ at sufficiently large $N$.

\begin{figure}
\begin{center}
$\begin{array}{l c r}
\includegraphics[width=10.2cm]{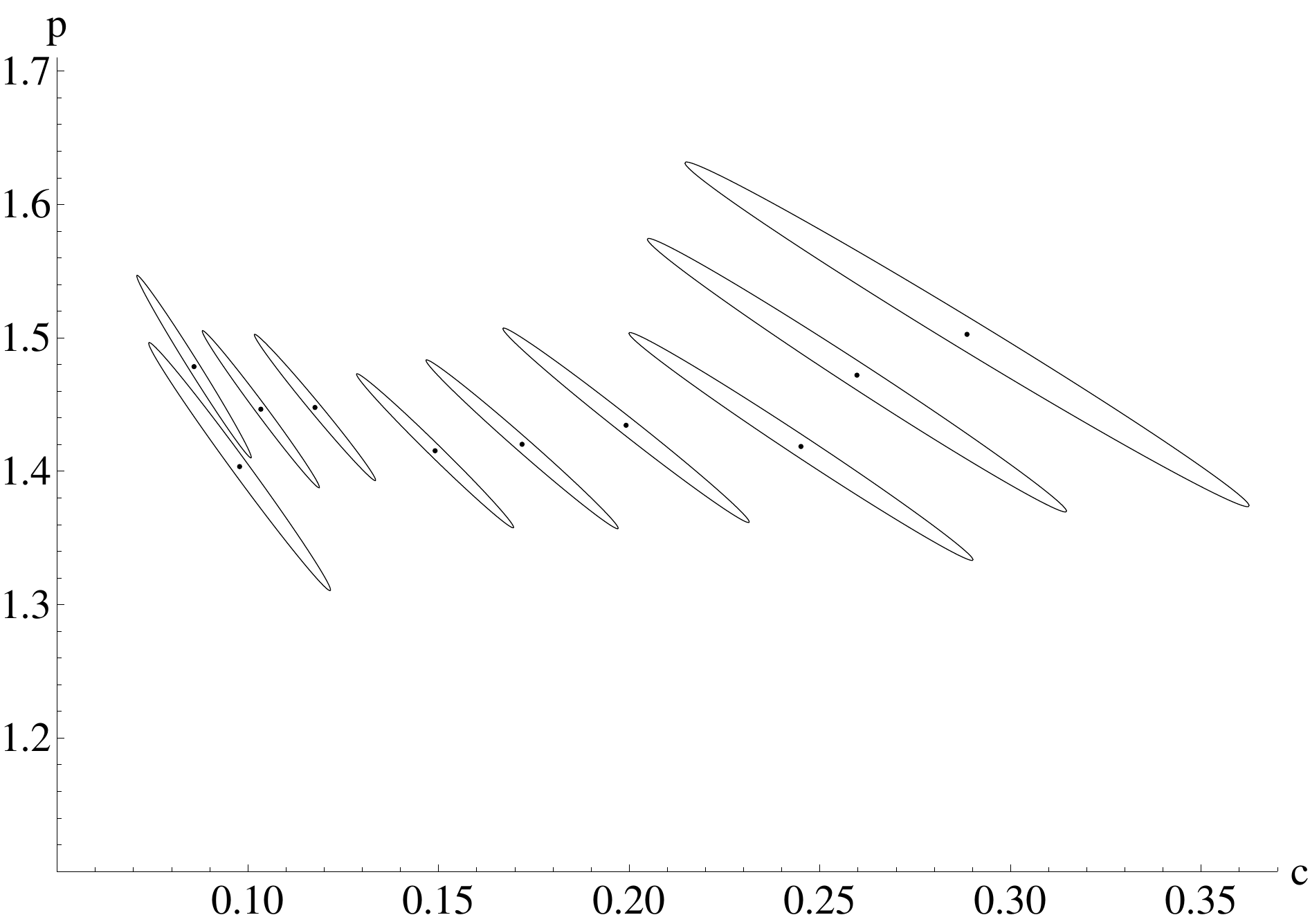} %%v2 1/10 figure axis label has been changed  %%v2 1/11 caption
\end{array}$
\caption{The ellipses show the 2$\sigma$ allowed regions of the $p-c$ plane, cf.~equation (\ref{eq:prob}), for $\omega=0.1,0.2,\ldots, 1.0$, from left to right, with $2 \le N \le 23$.
As $\omega$ increases (so that for fixed $F$ the cosmological constant decreases), $c$ increases substantially, while $p$ increases slightly.
%Thus, within this class of critical points, the probability of stability is higher at higher cosmological constant.
}
\label{fig_MMWpvsc}
\end{center}
\end{figure}

Next, Figure \ref{fig_MMWpvsc} shows the trends in $c$ and $p$, cf.~equation (\ref{eq:prob}), as $\omega$ is varied in $[0.1,1]$.  There is a distinct increase in $c$ as  $\omega$ increases, while $p$ shows a barely significant increase, so that overall the probability  of positivity  drops substantially as  $\omega$ increases.
This trend can be understood as follows: increasing $\omega$ at fixed $F$ reduces ${\cal H}_{\rm shift}$, and hence shifts the entire spectrum toward more negative values, making a fluctuation to positivity more improbable.

We note that for fixed $F$, increasing $\omega$ {\it{reduces}} the cosmological constant, so within this class of critical points,  the probability of stability is higher at higher cosmological constant.  One should not read too much into this, however, as one can increase the cosmological constant by increasing $F,|W|,m_{\rm susy}$ by a common factor without affecting the probability of stability.

\section{Stability of Critical Points with Small F-terms}
\label{sec:DD}

%%v2
The conclusions of \S\ref{sec:MMW} apply to a generic critical point, by which we mean one at which the functions $K, W$ are random functions that do not automatically manifest any special hierarchies.\footnote{Of course, large ratios can arise in this setting by chance, but this possibility is already encoded in the results of \S\ref{sec:MMW}.} However, it is far from clear a priori that a typical metastable vacuum arises from among the set of generic critical points:
a tiny subclass of critical points that enjoy a high likelihood of stability as a consequence of some special structure might well account for most of the metastable vacua.

As originally noted by Denef and Douglas in
\cite{DD}, a particularly interesting class of critical points are those at which the F-terms are small compared to the supersymmetric masses: approximate supersymmetry can be expected to make stability more likely.
Specifically, we will consider de Sitter critical points at which
 \be
 \sqrt{3}|W| <  F\ll |Z_{ab}| \sim |U_{abc}|% \in |R_{a \bb c\bd}|
 \, . \label{eq:DDscaling}
 \ee
In this section we will reexamine the stability of critical points in this approximately-supersymmetric corner of the supergravity landscape.
As will become clear, our conclusion differs from that of Denef and Douglas, and we will carefully explain
the reason for the disparity.

We will see that eigenvalue repulsion in the mass matrix between the bulk of the eigenvalues and the Goldstino direction typically generates at least two tachyonic directions, rendering generic critical points unstable.  This  effect significantly influences the fine-tuning needed to obtain a metastable de Sitter solution in supergravity.  Through numerical simulations and through statistical analysis we will show  that metastable critical points constitute an exponentially small fraction of all critical points: $N_{\rm vacua} \simeq N_{\rm crit. pts.} e^{-c N^p}$, with $p \approx 1.3$ and $c \approx 0.08$.

\subsection{The Denef-Douglas landscape of de Sitter vacua}
\label{subsec:DD}

In order to analyze the stability properties of the mass matrix in the regime where $F \ll m_{\rm susy}$,
we write ${\cal H}$ as
\cite{DD},
\be
{\cal H} =  V_0'' + V_1'' + V_2'' \, , \label{EqH}
\ee
with
\begin{equation}
 V_0'' = \left( {\cal{M}}  + |W| \mathbb{1} \right)\left({\cal{M}} - 2 |W| \mathbb{1} \right)\,,
\end{equation}
 \be
 V''_1 = \left(
 \begin{array}{c c}
 0 & S_1 \\
 \bar{S}_1 & 0
 \end{array}
\right) \, ,\   \    S_1 = U_{abc} \bF^c \, ,   \label{v1def}
 \ee
 and
 \be
 V''_2 =
 \left(
 \begin{array}{c c}
 S_2 & 0 \\
 0 & \bar{S}_2
 \end{array}
 \right) \, , \   \    S_2 = \delta_{a \bb}F^2  - F_a \bar{F}_{\bb} - R_{a \bb c \bar d} \bar{F}^c F^{\bar d}  \, ,
 \ee
where ${\cal{M}}$ is the matrix given in equation \eqref{eq:M}.

When $m_{\rm susy} \gg F$, all but two of the eigenvalues of ${\cal H}$ are generically of order $m_{\rm susy}^2$,
%%v2
and are predominantly determined by $V_0''$, with corrections of order $F$ from $V_1''$, and of order $F^2$ from  $V_2''$.
However, the critical point condition, equation \eqref{eq:M}, requires that ${\cal{M}}$ has an eigenvalue $2|W|$, with the corresponding eigenvector pointing in the Goldstino direction.
The eigenvalues of ${\cal{M}}$ come in pairs differing only in sign, and the eigenvalues $\lambda_{\pm} = \pm 2|W|$ of ${\cal{M}}$ correspond to eigenvalues $m_0^2 = 0$,  $m_0^2 =4|W|^2$
of $V_0''$.  The larger of these `Goldstino' eigenvalues is ${\cal O}(F^2)$, so that one cannot a priori neglect the effects of $V_1''$, $V_2''$ on the stability of the Goldstino direction.  This section is dedicated to a careful examination of these effects.

Setting $F_a = \delta_a^{~1} F e^{i \vartheta_F}$ and performing a unitary transformation that diagonalizes ${Z_a}^{\bc} \bar{Z}_{\bb\bc}$, we obtain
the simplified mass matrix
\bea
{\cal H}_{\star}
&=&
\left(
\begin{array}{l  l |  r r}
& m^2_{a \bb} & m^2_{a b} & \\ \hline
&m^2_{\ba \bb} & m^2_{\ba b}&
\end{array}
\right) \nonumber \\
&=&
\left(
\begin{array}{c | c}
\begin{array}{cc}
m^2_{1 \bar 1} & {\cal O}(F^2)  \\
{\cal O}(F^2) & {\rm diag}(\lambda_{a'}^2) \\
\end{array} & m^2_{ab}\\ \hline
m^2_{\bar a \bar b} &
\begin{array}{cc}
m^2_{1 \bar 1} & {\cal O}(F^2) \\
{\cal O}(F^2) & {\rm diag}(\lambda_{a'}^2 ) \\
\end{array} \\
\end{array}
\right) \,  . \label{Hess}
\eea
The eigenvalues of ${Z_a}^{\bc} \bar{Z}_{\bb\bc}$ have been denoted $\lambda^2_{a'}$, for $a'=2,\ldots N$, while
by the critical point equation we have
$m^2_{1 \bar 1} =   2 |W|^2- R_{1 \bar 1 1 \bar 1} F^2 $.
In the approximately-supersymmetric regime, $\lambda^2_{a'} \gg m_{\rm susy} F$, and we have correspondingly omitted ${\cal O}(F^2)$
%%v2 changed exponent
contributions to the diagonal entries
%%v2 1/10
$\left({\cal H}_{\star}\right)_{a'\bar a'}$ for $a'=2,\ldots N$.  The notation ${\cal H}_{\star}$ emphasizes that the matrix appearing in (\ref{Hess}) is not a truncation of ${\cal H}$ to some order in $F$.  Instead, ${\cal H}_{\star}$ has been strategically simplified so that, while it efficiently yields results for the two smallest eigenvalues of ${\cal H}$ that are accurate up to ${\cal O}(F^3)$ corrections, the higher eigenvalues of ${\cal H}_{\star}$, which are generically positive in any case, do not coincide with those of ${\cal H}$ to this accuracy.

Following the discussion of \cite{DD}, we focus on the submatrix spanned by the normalized eigenvectors of ${\cal{M}}$ with eigenvalues $\pm 2 |W|$, which in the above basis can be expressed as
\be
\big(\Psi^{ +}_{11}\big)_a = \frac{1}{\sqrt{2}} \big( e^{i \Delta \vartheta}\ \delta_a^{~1} + e^{- i \Delta \vartheta} \delta_a^{~N+1} \big)\   \   {\rm and}\    \big(\Psi^{ -}_{11}\big)_a = \frac{i}{\sqrt{2}} \big( e^{i \Delta \vartheta}\ \delta_a^{~1} - e^{- i \Delta \vartheta}\  \delta_a^{~N+1} \big) \, ,
\ee
where $ \Delta \vartheta =  \vartheta_F - \vartheta_W$.
Neglecting for the moment mixings with the other eigenvectors of ${\cal{M}}$, this $2\times2$ Goldstino submatrix of the full mass matrix is
%%v2
\bea
{\cal H}_{\rm sub}
=
\left(
\begin{array}{c c  }
m^2_{1 \bar 1}  & m^2_{11}  \\
 m^2_{\bar 1 \bar 1}& m^2_{1 \bar 1}
 \end{array}
\right) \,  .
\eea
While a diagonalization of this subsystem by itself does {\it{not}} in general correspond to a diagonalization of the corresponding directions in the full mass matrix, it is instructive to attempt to treat the off-diagonal mixings in perturbation theory. The eigenvalues of the submatrix are given by
\be
h^{\pm} = m^2_{1 \bar 1} \pm |m^2_{11}| \, , \label{lambdapm}
\ee
which can be written as
\be
h^{\pm} = 2 |W|^2 - R_{1 \bar 1 1 \bar 1} F^2 \pm \Big| U_{111} F e^{-\vartheta_F} - 2 |W|^2 e^{2 i ( \vartheta_F - \vartheta_W)} \Big| \, .
\ee
The dominant contribution to $h^{\pm}$ for $|U_{111}| \sim  |R_{1 \bar 1 1 \bar 1}| \sim {\cal{O}}(F^0)$ is the term $U_{111} F$, so that generically $h^-<0$.

In \cite{DD}, it was observed that fine-tuning $|U_{111}|$ to be ${\cal{O}}(F)$ is necessary  for stability of the mass matrix. However, \cite{DD} also argued that this condition is sufficient, and concluded that metastable critical points are fairly common in a supergravity landscape.
We will now show that  the eigenvalues $h^{\pm}$ of the submatrix
%%v2
${\cal H}_{\rm sub}$ of ${\cal H}$ cannot be regarded as good approximations to the actual eigenvalues of ${\cal H}$.  Upon computing the leading-order corrections to \eqref{lambdapm}, we will find that metastable critical points constitute an exponentially small fraction of all de Sitter critical points.

\subsection{Eigenvalue repulsion induces tachyons}
In this section we discuss the correction to the eigenvalues of the mass matrix induced by
$V_{1}^{''}$, equation (\ref{v1def}).
To determine the two smallest eigenvalues to ${\cal O}(F^2)$, we may neglect
${\cal O}(F^2)$ contributions to
%%v2 1/10 $m^2_{a' \bb'}$ for $a' \neq \bb'$.
$m^2_{a' \bb'}$ for $a' \neq b'$.
With this simplification,
equation (\ref{Hess}) can be written as
\be
{\cal H}_{\star} \simeq
\left(
\begin{array}{c c c | c c c}
m^2_{1 \bar 1} & 0&  & s & v^T_{a'}    \\
0 & {\rm diag}(\lambda_{a'}^2) & & v_{a'} & T_{a' b'}  \\ \hline
s^* & v^{\dagger}_{\ba'} &&  m^2_{1 \bar 1} & 0 & \\
 v^*_{\ba'} & T^{*}_{\ba' \bb'} & & 0 & {\rm diag}(\lambda_{a'}^2 ) & \\
\end{array}
\right) \,.  \label{Hsimple}
\ee
In equation \eqref{Hsimple} we have introduced the $U(N-1)$ scalar $s = m^2_{11}$,  the vector $v_{a'} = m^2_{1 a'}$, and the symmetric tensor $T_{a' b'} = m^2_{a' b'}$. The unitary transformation
\bea
U =
\left(
\begin{array}{c c c | c c c}
- \frac{e^{i \alpha} }{\sqrt{2}} & 0 & & \frac{e^{i \alpha} }{\sqrt{2}} & 0 \\
0 & \delta_{a' \bb'}  & & 0 & 0 \\ \hline
\frac{1}{\sqrt{2}} & 0 & & \frac{1}{\sqrt{2}} & 0 & \\
0 & 0 & & 0 & \delta_{a' \bb'}  \\
\end{array}
\right) \, ,
\eea
with $ \alpha = {\rm arg}(s)$, diagonalizes the sub-matrix ${\cal{H}}_{\rm sub}$, i.e.\
\bea
\widetilde{{\cal H}}_{\star} = U^{\dagger} {\cal H}_{\star} U =
\left(
\begin{array}{c c c | c c c}
h^{-} & u^{\dagger}_{a'} && 0 & w^T_{a'} \\
u_{a'} & {\rm diag}(\lambda^2_{a'}) && u_{a'} & T_{a' b'} \\ \hline
0 & u^{\dagger}_{\ba'} & & h^{+} & -w_{a'} \\
w^*_{a'} & T^*_{\ba' \bb'} && -w^*_{a'} & {\rm diag}(\lambda^2_{a'} )\\
\end{array}
\right) \, , \label{eq:Htilde}
\eea
where $u_{a'} = \frac{1}{\sqrt{2}} v_{a'}$, $w_{a'} = - \frac{1}{\sqrt{2}} e^{-i \alpha}  v_{a'}$, and $h^{\pm}$ is given by equation \eqref{lambdapm}. The conclusion of \cite{DD} is that modest fine-tuning of $s$ ensures the positivity of $h^{\pm}$,
and hence of ${\cal H}$.
Here we investigate the effect of the vector $v_{a'}$  on the eigenvalues of ${\cal H}$.  The leading-order effect of the tensor $T_{a'b'}$  is to induce ${\cal O}(m_{\rm susy} F)$ shifts of the eigenvalues $\lambda^2_{a'} \gg m_{\rm susy} F$, so that we may consistently neglect $T_{a'b'}$.  The characteristic polynomial of $\widetilde{{\cal H}}_{\star}$ is then given by
\bea
{\cal C}(\rho) &=& {\cal C}_0(\rho) \Biggl[1 - \sum_{b' =2}^{N} \frac{|v_{b'}|^2 }{(h^+ - \rho)(\lambda^2_{b'} - \rho)}  - \sum_{b'=2}^N \frac{|v_{b'}|^2  }{(h^- - \rho)(\lambda^2_{b'} - \rho)} \nonumber \\
&+& \sum_{a', b' = 2}^N \frac{|v_{a'}|^2 |v_{b'}|^2 }{(h^+ - \rho)(h^- - \rho)(\lambda^2_{a'} - \rho)(\lambda^2_{b'} - \rho)} \Biggr]
\, , \label{characteristic}
\eea
where ${\cal C}_0(\rho)$ denotes the characteristic polynomial for $v_{a'}=0$, i.e.
%%v2
\be
{\cal C}_0(\rho)= (h^- - \rho)(h^+ - \rho)\prod_{a'=2}^N \big(\lambda^2_{a'} - \rho \big)^2\,.
\ee

The leading-order effect of the vector $v_{a'}$ is evidently to induce an `interaction' between the approximate eigenvalues $h^{\pm}$ and $\lambda^2_{a'}$. This interaction is a manifestation of eigenvalue repulsion, and indeed, the effect of each term in the sum is to increase the splitting between $h^{\pm}$ and $\lambda^2_{a'}$. Restricting the polynomial to small values of $\rho$ close to the smallest eigenvalues of the mass matrix and dividing by the overall factors of the larger eigenvalues, equation (\ref{characteristic})
can be rewritten as
\bea
\frac{{\cal C}(\rho)}{\prod_{a'=2}^N \lambda^2_{a'}} &=& (h^+ - \rho)(h^- - \rho) - (h^- - \rho) \sum_{b'=2}^N \frac{|v_{b'}|^2}{\lambda^2_{b'}} - (h^+ - \rho) \sum_{b'=2}^N \frac{|v_{b'}|^2}{\lambda^2_{b'}} \nonumber \\
&+&   \sum_{a',\ b' =2}^N \frac{|v_{a'}|^2\ |v_{b'}|^2}{\lambda^2_{a'} \lambda^2_{b'}} \, .
\eea
The smallest eigenvalues of the mass matrix are thus given by
\bea
m^2_{\pm} &=& h^{\pm} -   \sum_{b'=2}^N \frac{|v_{b'}|^2}{\lambda^2_{b'}} = m^2_{1 \bar 1} \pm |m^2_{11}| -   \sum_{b'=2}^N \frac{|m^2_{1 b'}|^2}{\lambda^2_{b'}} \nonumber \\
&=&  2 |W|^2 + K_{1 1}^{~~e} K_{\bar 1 \bar 1  e} F^2- K_{1 \bar 1 1 \bar 1} F^2 \pm \Big|  U_{111} F e^{-\vartheta_F} - 2 |W|^2 e^{2 i ( \vartheta_F - \vartheta_W)} \Big| \nonumber \\
&-&   F^2 \sum_{b'=2}^N \frac{|U_{11b'}|^2}{\lambda^2_{b'}}   \, . \label{eq:Meig}
\eea
This is one of our main results.  The smallest eigenvalue $m^2_-$ of the Hessian matrix ${\cal H}$ differs from that of \cite{DD} by the non-positive term
$ -  F^2 \sum_{b'=2}^N \frac{|U_{11b'}|^2}{\lambda^2_{b'}}$, in a manifestation of eigenvalue repulsion between the Goldstino and the supersymmetrically stabilized moduli with masses of order $\lambda^2_{b'}$.

We now turn to assessing the impact of this contribution to $m^2_-$.

\subsection{Eigenvalue fluctuations and de Sitter vacua}
\label{sec:DDdS}

In this section we will determine the probability that a randomly chosen approximately-supersymmetric critical point is metastable by computing the probability that $m^2_-$, as given in equation \eqref{eq:Meig}, is positive.
Our approach is to determine the statistical properties\footnote{Recall from \cite{DD} that a fine-tuning of $ |U_{111} F| \lesssim F^2$ is necessary for stability, and granting this fine-tuning, all the terms in equation \eqref{eq:Meig} are of the same order, ${\cal O}(F^2)$.} of each term in the sum, i.e.\ we will obtain the cumulative distribution function (cdf) for each term, from which the corresponding probability density function (pdf) can be obtained by differentiation.  Although in principle one might hope to convolve the constituent probability density functions to obtain the pdf of $m^2_-$, this is rather involved.
Fortunately, we will find that one of the terms of equation \eqref{eq:Meig} dominates both in magnitude and in the probability of fluctuations, and it suffices to examine this term.

For the analytical estimates provided here, we will assume $\Omega = {\cal N}(0,{ \scriptstyle \frac{1}{\sqrt{N} } })$, though similar arguments could be made for e.g.\ the uniform distribution.

We find it convenient to rewrite \eqref{eq:Meig} as
\be
 m^2_- = F^2 {\cal{T}} + F^2 {\cal{S}} \label{eq:DDmm}
\ee with
\begin{equation}
{\cal{T}} = \frac{2}{3}\omega^2 +  K_{1 1}^{~~e} K_{\bar 1 \bar 1  e}   - K_{1 \bar 1 1 \bar 1}   - |t_{\rm hol}| \, , \label{eq:DDmmT}
\end{equation} and
\begin{equation}
{\cal S} \equiv -  \sum_{b'=2}^N \frac{|U_{11b'}|^2}{\lambda^2_{b'}}\,,  \label{sis}
\end{equation}
where
\begin{equation}
|t_{\rm hol}| \equiv \Big| U_{111} e^{2 i \vartheta_W - 3 i \vartheta_F} F^{-1}-\frac{2}{3}\omega^2  \Big|\,,
\end{equation} and we have used the definition (\ref{defineomega}).

\subsubsection{Subdominant contributions}\label{unperturbed}

We will begin by studying the terms collected in ${\cal T}$,
which do not involve the vector $v_{a'}$.

At the critical points of interest, $\sqrt{3}|W|\le F$, so that the total energy density is nonnegative.  Thus, the first term in equation \eqref{eq:DDmmT} gives a contribution in the range $[0, \frac{2}{3}]$.

The second term of equation \eqref{eq:DDmmT} is $|K^{(3)}|^2 \equiv K_{1 1}^{~~e} K_{\bar 1 \bar 1  e}$, which is the sum of squares of $N$ random variables, each drawn from ${\cal N}(0,{ \scriptstyle \frac{1}{\sqrt{N} } })$.
Thus, $|K^{(3)}|^2$ is distributed
%%v2
as $\frac{1}{N}\chi^2_N$, where $\chi^2_N$ is a
%according to a
chi-square distribution with $N$ degrees of freedom.  Since $\chi^2_N$ has mean $N$, we conclude that
\begin{equation}
\langle\ | K^{(3)}|^2\ \rangle     =1 \,.
\end{equation}
To find the probability of fluctuations, we note that
the corresponding cdf is given by
\be
P\Bigl(| K^{(3)}|^2\leq x\Bigr) = P\Bigl(\frac{1}{N} \chi^2_N \leq x\Bigr)=  \frac{1}{\Gamma(N/2)} \gamma \Big(\frac{N}{2}, \frac{N x}{2}\Big)\,,
\ee
where $\gamma$ denotes the  lower incomplete gamma function. The asymptotic behavior can be obtained as follows: by the central limit theorem, the cdf of a chi-square distributed variable for
$N \gg 1$ degrees of freedom tends to that of a Gaussian distributed variable with unit variance,
$
 P(\chi^2_{N} \leq  y) \approx  P({\cal N}(0,1) \leq x) \,,
$
where $x = \frac{y - N}{\sqrt{2N}}$ \cite{Abramowitz}.

We are particularly interested in the probability of $|K^{(3)}|^2$ fluctuating to a large value and thereby stabilizing the smallest eigenvalue $m^2_-$ of ${\cal H}$. As we will describe below, large in this context means ${\cal O}(N)$, so that we consider
\be
P\Bigl(|K^{(3)}|^2 \leq N \Bigr) \approx  P\Bigl({\cal N}(0,1) \leq 2^{-1/2}N^{3/2}\Bigr) \, ,
\ee
for $N\gg1$, from which we obtain
\be
P\Bigl(|K^{(3)}|^2 \geq N \Bigr) \lesssim \frac{1}{\sqrt{ \pi} N^{3/2}} \  e^{- \frac{N^3}{4}}\, .
\ee  Fluctuations of $|K^{(3)}|^2$ may therefore be neglected in comparison to the much more probable fluctuations we will discuss in \S\ref{ert}.

The third term in \eqref{eq:DDmmT}, $ K^{(4)}_{1 \bar 1 1 \bar 1} $, is normally distributed with a vanishing expectation value, and with a variance no larger than $\frac{2}{N}$, so that large deviations of the order $N$ are likewise so improbable as to be negligible:
\be
P\Bigl(K^{(4)}_{1 \bar 1 1 \bar 1} \geq N\Bigr) \sim e^{-N^3} \, .
\ee

The fourth term in \eqref{eq:DDmmT}, $-|t_{\rm hol}|$, is negative semidefinite, and only one {\it{entry}} (not eigenvalue) of ${\cal H}$, namely $U_{111}$, needs to be adjusted in order to change the size of $|t_{\rm hol}|$.
Therefore, it is straightforward to fine-tune $|t_{\rm hol}|$ to be small.  It is clear from the discussion above that, as originally noted in \cite{DD}, $m^2_-$ is generically negative unless $U_{111}$ is fine-tuned to make $|t_{\rm  hol}| \lesssim {\cal O}(1)$.  For our goal of obtaining a conservative estimate of the probability that $m^2_->0$, it suffices to set $|t_{\rm hol}| =0$.

\subsubsection{The eigenvalue repulsion term}  \label{ert}

Finally, the last term in equation \eqref{eq:DDmm} is the sum of squares of $N-1$ terms.
The numerators of the terms in equation \eqref{sis} are the squares of independent normally distributed variables, while the denominators are the squares of the eigenvalues of ${\cal M}$.

The eigenvalues of ${\cal M}$ range from around  ${\cal{O}}(\frac{1}{N})$ to  $2$ in units\footnote{Since by assumption $U_{a b c} \sim Z_{ab}$, the dependence on the supersymmetric mass scale $m_{\rm susy}$ cancels between the numerator and denominator in equation (\ref{sis}).} of $m_{\rm susy}$,
so $\langle|{\cal S}|\rangle \sim {\cal{O}}(N)$.
Recalling that the contributions to ${\cal T}$
have mean sizes independent of $N$, we conclude that ${\cal S}$ provides the dominant contribution to $m^2_-$ at large $N$.
Since $\langle {\cal S}\rangle <0$, this term
{\it{destabilizes generic  critical points in the approximately-supersymmetric regime.}}

To determine the (small) probability that $m^2_-$ is nevertheless positive, we will now estimate the probability that ${\cal S}\gtrsim -1$, so that ${\cal T} + {\cal S}$ can be positive.
First, we recognize that in light of the discussion in \S\ref{sec:RMT}, fluctuations that increase the denominators appearing in ${\cal S}$, corresponding to inward fluctuations of the eigenvalues of a  Wishart matrix, are extremely unlikely at large $N$ \cite{KC}.
Fluctuations of ${\cal S}$ toward smaller magnitude are principally determined by the fluctuations of the numerators. (We have explicitly verified this in simulations.)
This justifies simplifying the problem by fixing the factors of  $ \lambda^2_{b'}$ to their mean values,  $\langle \lambda^2_{b'} \rangle$, as determined by the bulk distribution given by equation \eqref{wishs}.
Henceforth we consider the sum %%v2 \footnote{Please note the difference in sign between ${\cal S}$ and ${\cal S'}$.}
\be
   {\cal S'} = \sum_{b'=2}^N  \frac{|U_{11b'} |^2}{\langle \lambda^2_{b'} \rangle }  \, , \label{eq:Sprime}
\ee
which is the weighted sum of $N-1$ variables that are all independently distributed as $\chi^2_1$.  Weighted sums of $\chi^2$-distributed  variables (or, equivalently, sums of $\Gamma$-distributed variables with different scale parameters) occur frequently in statistics, and in particular in the theory of  the distributions of quadratic forms. While we have not found a closed-form expression for the convolution of $N-1$ such terms, approximations for expressions like \eqref{eq:Sprime} have been developed.  An approximation by  Solomon and Stephens has been argued to be particularly accurate in the small-argument regime of interest \cite{Solomonson},
%%v2 however
but we will find that for our purposes it does not constitute a close approximation to the cumulative probability for small arguments. By matching the first three algebraic moments $\mu_1$, $\mu_{2}$, $\mu_{3}$ of ${\cal S'}$ to those of  $a \cdot w^b$, where $w$ is  $\chi^2_r$-distributed and  $a, b,$ and $r$ are constants, this approximation is obtained\footnote{There is a misprint in the fifth equation of \S 3.1 of \cite{Solomonson}.} by numerically
%%v2
solving the equations
\bea
\mu_1 &=& a\ 2^b\ \frac{\Gamma(b+\frac{r}{2})}{\Gamma(\frac{r}{2})} \, , \label{eq:Approx1} \\
\frac{\mu'_2}{\mu_1^2} &=& \Gamma\Big(\frac{r}{2}\Big)\ \frac{\Gamma(2 b+\frac{r}{2})}{\left[ \Gamma(b+ \frac{r}{2}) \right]^2} \, , \label{eq:Approx2} \\
\frac{\mu'_3}{\mu_1^3} &=& \Big[ \Gamma\Big(\frac{r}{2}\Big) \Big]^2\ \frac{\Gamma(3 b+\frac{r}{2})}{\left[ \Gamma(b+ \frac{r}{2}) \right]^3}  \, . \label{eq:Approx3}
\eea
Thus, in this approximation,
\be
P\Bigl({\cal S'} \leq s\Bigr) \approx P\Bigl(a (\chi^2_r)^b \leq s\Bigr) = P\Bigl(\chi^2_r \leq  \Big( \frac{s}{a} \Big)^{1/b} \Bigr) \, .
\ee
As ${\cal T} \sim {\cal{O}}(1)$, $m^2_-$ could be positive if  ${\cal S'}$ fluctuates down to be ${\cal O}(1)$, for which we obtain
\be
 P\Bigl(m^2_->0\Bigr) \approx P\Bigl({\cal S'} \lesssim 1\Bigr) \approx e^{-c\cdot N^p} \, ,
 \ee
where $c\simeq 23$, and $p \simeq 0.24$. As we will see in \S\ref{sec:DDnum}, even though this approximation qualitatively matches the shape of ${\cal S'}$, for $N \gg 1$ it severely overestimates the probability of a fluctuation of ${\cal S'}$ to be of ${\cal O}(1)$, and it remains an open question to obtain a good analytic or semi-analytic approximation of equation \eqref{eq:Sprime}.

\begin{figure}
\begin{center}
$\begin{array}{l c r}
\includegraphics[width=10.2cm]{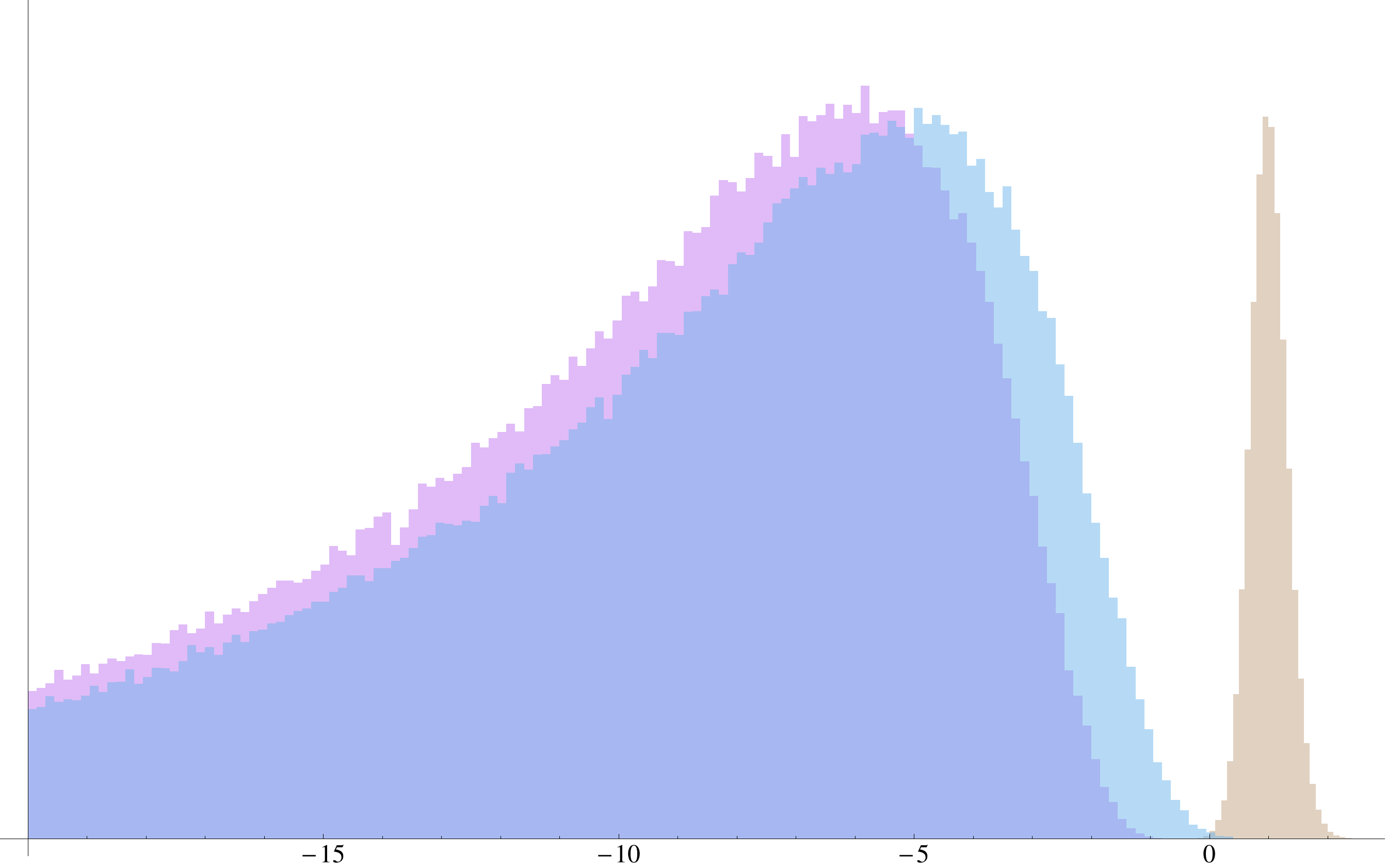}  %%v2 1/10 caption includes units
\end{array}$
\caption{Histograms of the smallest eigenvalue $m^2_{-}$, cf.\ equation (\ref{eq:DDmm}), and its constituent terms ${\cal T}$ and ${\cal S}$, for $N=40$, in units of $F^2$.
The eigenvalue repulsion sum ${\cal S}$ has the leftmost peak, the total mass $m^2_{-}$ has the central peak, and ${\cal T}$ appears on the right.
Note that ${\cal S}$, and consequently $m^2_{-}$, has support over a range of size $N$ (not fully shown in the figure), while ${\cal T}$ has variance $\frac{2}{N}$.
}
\label{fig_DDsum}
\end{center}
\end{figure}

\subsection{Numerical results}
\label{sec:DDnum}

Figure \ref{fig_DDsum} shows a histogram of $m^2_{-}$ and its constituent terms ${\cal T}$ and ${\cal S}$, for $N=40$.  It is clear that  ${\cal S}$ gives the dominant contribution to $m^2_{-}$.
Moreover, the narrow support of the  ${\cal T}$ histogram illustrates the finding of \S\ref{unperturbed} that large fluctuations of ${\cal T}$ are much less probable than correspondingly large fluctuations of ${\cal S}$.

\begin{figure}
\begin{center}
$\begin{array}{l c r}
\includegraphics[width=12.2cm]{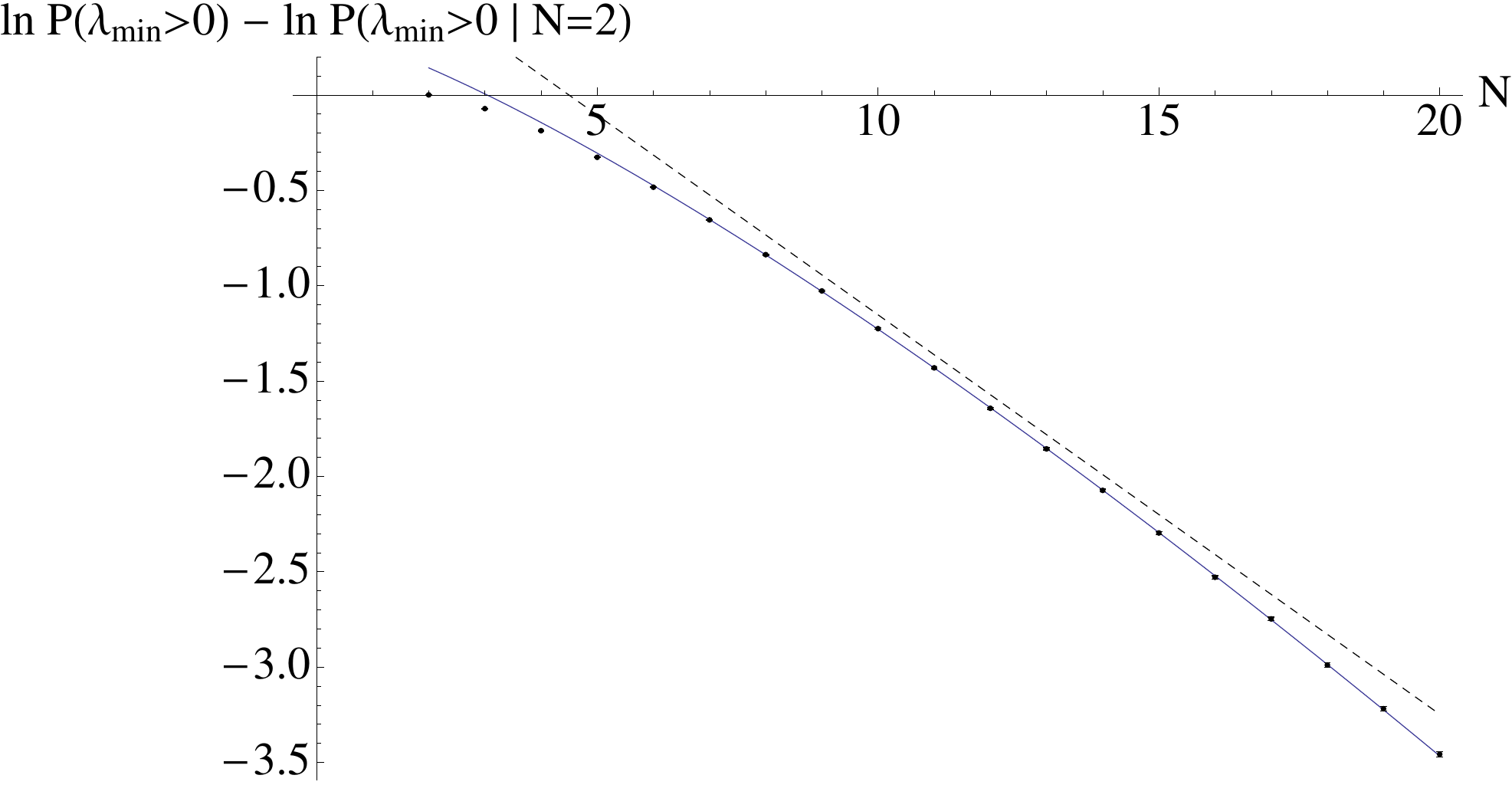}  %%v2 1/10 figure has been changed
\end{array}$
\caption{The logarithm of the probability $P(\lambda_{\rm min}>0)$ that the smallest eigenvalue of ${\cal H}$ is positive, as a function of $N$.
Each point corresponds to $10^6$ realizations of the full mass matrix, and the error bars give the 2$\sigma$ statistical uncertainty.
The curve shows the best fit to equation (\ref{eq:prob}), with $p=1.28 \pm 0.03$ and $c= 0.083 \pm 0.008$.
The dashed line with $p=1$ is for reference.}%%v2 min --> \rm min
\label{fig_DDN}
\end{center}
\end{figure}

Finally, Figure \ref{fig_DDN} presents the result of simulations of the mass matrix in the approximately-supersymmetric regime.  (The value of $\omega$ has a negligible effect on stability in this regime.)
The data agrees well with (\ref{eq:prob}), with $p=1.28 \pm 0.03$ and $c= 0.083 \pm 0.008$.\footnote{To obtain a conservative bound, we fit to the data points with $N \geq 7$.} %%v2
This is a much larger value for $p$ than that obtained by the analytical estimate of \S\ref{ert}, so that the latter gives an extremely conservative upper bound on the asymptotic large $N$ probability of positivity.

\section{Beyond Random Supergravity}\label{sec:KKLT}

In this section we will discuss potential extensions of our assumptions (\S\ref{assumptions}), explain the consequences of decoupling for the probability of positivity (\S\ref{sec:decoupling}), and illustrate our results in the example of the KKLT scenario (\S\ref{sec:KKLTactual}).

\subsection{Universality} \label{universalityintro}  \label{assumptions}

The results of \S\ref{sec:MMW} and \S\ref{sec:DD} were obtained from the assumptions enumerated in \S\ref{sec:RS}: most notably, $K$ and $W$ were taken to be random functions of $N$ scalar fields, so that their various (appropriately covariant) derivatives are i.i.d.\ variables drawn from a distribution $\Omega(\mu,\sigma)$.
Equivalent assumptions are standard in the statistical study of flux compactifications, cf.\ \cite{Douglas:2006es}, and in particular are fully consistent with the assumptions of \cite{DD}.
Nevertheless, in this section we will venture a few remarks about possible extensions of this simplest definition of a random supergravity.

First, we have taken the random variables to be normally distributed, $\Omega = {\cal{N}}(0,\frac{1}{\sqrt{N}})$, throughout this work, and now we justify this assumption.
The celebrated phenomenon of universality in random matrix theory ensures that at large $N$ the eigenvalue spectrum, and also the fluctuations of extreme eigenvalues, are independent of the statistical details of the inputs.
Universality has been demonstrated in a staggering array of physical and mathematical systems, many with $N \lesssim {\cal O}(10^2)$, including interfaces in liquid crystals \cite{lc}, the timing of buses in Cuernavaca \cite{bus}, and the power output of coupled lasers \cite{lasers}.  See \cite{Deift, KUniv} for overviews of universality and \cite{Bai, Soshnikov} for results with close connections to ensembles studied here;  extensions to ensembles in which the matrix entries have power-law tails include \cite{heavy}.  The lesson is that the particular choice of $\Omega$ is immaterial, provided that the moments of $\Omega$ are appropriately bounded.  (One should compare distributions $\Omega_{1}$, $\Omega_{2}$ yielding the same root-mean-square size for entries in ${\cal H}$, as this sets the physical scale.)

Despite the strong expectation that universality should be applicable for our system, it is still reasonable to ask whether the values of $N$ in our analysis are large enough for these asymptotic results to apply in practice.  We have addressed this point directly by repeating our simulations for different choices of distribution, with excellent agreement.

A more fundamental question is whether in the effective theories derived from string compactifications, the derivatives of $K$ and $W$ are accurately modeled as i.i.d.\ variables drawn from {\it{any}} distribution, or if instead these quantities are not i.i.d.\footnote{We thank M.~Douglas for instructive correspondence on this point.}  Microphysical constraints, for example the relics of extended supersymmetry,\footnote{See \cite{Borghese:2011en} for related work in maximal supergravity.}
 might be expected to introduce correlations among these variables, as in the special geometry relation of (\ref{sg}), cf.~\cite{DD,Covi}, so that the derivatives of $K$ and $W$ are not all independent.   A definitive answer to this question is beyond the scope of this work, but it is encouraging that in simpler cases such as the Wigner ensemble, universality has been shown to apply to matrices with highly correlated entries \cite{SSB}.

% \subsubsection{Global constraints}
It would also be interesting to understand the possible impact of global constraints on our considerations.
We have studied the ensemble of critical points arising in a general supergravity theory, and we have taken the derivatives of $K$ and $W$ evaluated at each such point to be random functions.  To understand the distribution of vacua within the moduli space, one should incorporate further structure.  The (index) density of supersymmetric vacua is well known to be correlated with the curvature of the moduli space \cite{AshD}, while the global structure of the superpotential is better modeled as a random holomorphic section of a line bundle over the moduli space (see \cite{DSZ} for a definitive treatment of the density of supersymmetric vacua in this context).  Extending the study of non-supersymmetric vacua to this level of detail is an interesting problem for the future.

%%v2
One might expect that constraints from Morse theory will require some small deviations from the purely statistical results obtained here.\footnote{We thank B.~Czech for very helpful correspondence about these constraints.}  The random matrix ensembles we have described (as in \cite{DD} and earlier works) predict certain ratios between the numbers of saddle points of varying index, which are not automatically consistent with the Morse inequalities.  The necessary adjustments can be accommodated without changing the number of minima relative to saddle points, and we find it plausible that any effect on the relative number of minima can be neglected.

\subsection{Decoupling improves stability}\label{sec:decoupling}

A significant assumption in our analysis is that $W$ and $K$ are general random functions of $N$ scalars.   In physically well-motivated examples, there can of course be two or more sectors of fields with distinct mass scales.
For instance, consider\footnote{We are indebted to S.~Kachru for emphasizing the importance of this example.} a two-sector supergravity theory with $N=N_H+N_L$ fields, in which the heavy scalars $\phi_{H}^a$, $a=1,\ldots N_H$,  receive large supersymmetric masses $m_H$, and supersymmetry is dynamically broken in a decoupled system of lighter scalars  $\phi_{L}^{i}$, $i=1,\ldots N_L$, at a much lower scale $m_L$.

Explicitly, such a model can be constructed from a superpotential and a K\"ahler potential that are additively separable. In a convenient K\"ahler gauge, one has
\bea
K(\phi_{H}, \bar \phi_{H}, \phi_L, \bar \phi_L) &=&   K_H(\phi_{H}, \bar \phi_{H}) +  K_L(\phi_L, \bar \phi_L) \, , \nonumber \\
W(\phi_H, \phi_L) &=&   W_H(\phi_H) +   W_L(\phi_L) \, . \label{eq:WKsep}
\eea
By assumption $Z_{a b} \sim m_H$ and $Z_{ ij } \sim m_L$, while  by \eqref{eq:WKsep}, the cross-couplings in the supersymmetric mass matrix are small: $Z_{ai} =   K_{a} W_i \sim  {\cal{O}}(F)$. Thus, at small $F$, ${\cal H}_{\rm susy}$ separates into two distinct Wishart matrices. (If $F$ is not small compared to $m_L$, or if the separability of the superpotential is imperfect, then  the off-diagonal masses in ${\cal H}_{\rm susy}$ cannot be neglected.)

A cautionary remark is necessary at this point.  The masses-squared in a supersymmetrically-stabilized sector are not necessarily positive: setting $F=0$ in equation (\ref{MasterHessian}), the contribution of ${\cal H}_{\rm susy}$ is nonnegative, but  ${\cal H}_{\rm shift}$ and ${\cal H}_{\rm pure}$ make tachyonic contributions that will be significant unless $W \ll m_{\rm susy}$.  Of course, the resulting masses do obey the Breitenlohner-Freedman (BF) bound, but this in itself  does not guarantee that this sector will remain stable after supersymmetry is broken badly enough to make the cosmological constant positive. % The lesson is that to guarantee the stability of a sector of supersymmetrically-stabilized fields, one should ensure that $m_{\rm susy} \gg W$, so that ${\cal H} \approx {\cal H}_{\rm susy}$.

We now recall from \S\ref{sec:DD} that for $F\ll m_{\rm susy}$,
%We have emphasized that in the approximately-supersymmetric regime in which the F-terms are small compared to the supersymmetric masses,
superpotential couplings of the form $\frac{|U_{11A'}|^2}{\lambda^2_{A'}}$ contribute to the destabilization of the Goldstino direction, cf.~equation \eqref{eq:Meig}, where $A'$ runs over all fields.
%Here $a'$ run over all directions perpendicular to the complex Goldstino direction, which in a supergravity arising as a dimensional reduction of a compactification of string theory correspond to the moduli.
%The tachyonic instability in the Goldstino direction is sensitive to the number of fields that couple to the Goldstino through $U_{11 A'}$, where $A'$ runs over all fields in both sectors.
The numerator of this contribution to the Goldstino mass from the heavy, supersymmetric subsystem is
 \be
 |U_{1 1 a}|^2  = |{\cal D}_{a} Z_{1 1}|^2 = |\partial_a Z_{1 1} + K_a  Z_{1  1} |^2 %\sim
 %\partial_{\phi^1}\Big( {\cal D}_{\phi^1} F_i \Big) - \Gamma^{\phi^a}_{\phi^1 \phi^1}\ {\cal D}_{\phi^a} F_i
 %m^2_L
  \, ,
 \ee
which under the decoupling assumptions of equation \eqref{eq:WKsep} is of order $|W|^2$.  (For a non-decoupled system one finds instead $|U_{11a}|^2 \sim m_{\rm susy}^2$.)

Since the denominator $\lambda^2_{a}$ is of order $m_H^2$, the separability of equations \eqref{eq:WKsep} leads to a suppression of order $\frac{|W|^2}{m^2_H}$
of the heavy fields' negative contribution to the Goldstino direction mass-squared.  Thus, even for a modest hierarchy between the supersymmetric masses, $m_H^2  \gtrsim N_H |W|^2$, the high-scale sector decouples,
and does not contribute significantly to the mass of the Goldstino.

In conclusion, the relevant number of fields for the stability analysis of \S\ref{sec:DD} is $N_L$, the number of `light' fields that participate in dynamical supersymmetry breaking (the  superpartner of the Goldstino is assumed to be entirely among these fields.)  The fraction of critical points that are metastable is then proportional to  ${\rm exp}(-c {N_L}^p)$.  Provided that the heavy sector, taken in isolation, contains a number of supersymmetric vacua that is exponential in $N_H$, then the net result, for $N_H \gg N_L$, is a mild reduction in the number of metastable vacua.

\subsection{Stability in the KKLT scenario} \label{sec:KKLTactual}

The KKLT scenario \cite{KKLT} provides a useful setting to illustrate our findings.  Consider a model with $N_K \equiv h^{1,1}_{(+)}$ K\"ahler moduli $T_i$, $i=1,\ldots N_K$, and $N_C \equiv h^{2,1}$ complex structure moduli $\zeta_a$, $a=1,\ldots N_C$.
Suppose that the superpotential takes the form
\begin{equation} \label{kkltw}
W = \int G\wedge\Omega + \sum_{i=1}^{N_K} {\cal{A}}_{i}(\zeta)\ {\rm{exp}}\left(\frac{2\pi}{n_{i}}T_i\right)\,,
\end{equation} where $n_i$ is the dual Coxeter number for superpotential terms generated by gaugino condensation, $n_i=1$ for terms generated by Euclidean D3-branes, and $\int G\wedge\Omega$ depends on the $\zeta_a$.
Finding a compactification with many moduli for which each K\"ahler modulus appears in the nonperturbative superpotential is a difficult task  (cf.\ \cite{DDF, DDFGK} for detailed examples).  Our purpose is to show that, {\it{granting}} a superpotential of the form (\ref{kkltw}), then for $N_K \gg 1$, an exponentially small fraction of de Sitter critical points are metastable vacua.

An important scale in the problem is the flux scale $m_{\rm flux}$, which sets the typical size of the supersymmetric masses for the $\zeta_a$.  In light of the very large number of choices of quantized flux \cite{Bousso:2000xa}, one can find configurations in which the vacuum expectation value of the classical superpotential obeys $\langle  \int G\wedge\Omega \rangle \ll m_{\rm flux}^3 $.
This fine-tuning is necessary in order to obtain a parametrically controlled vacuum with a reasonably small cosmological constant.  Given such a flux superpotential, one can find \cite{KKLT} a supersymmetric AdS vacuum with all moduli stabilized.

Our goal is to assess the stability of such a configuration after uplifting to de Sitter space.  As a conservative first step, we imagine that the uplifting increases the cosmological constant without creating new instabilities, as a (fictitious) moduli-independent D-term would do.  We expect that more plausible sources of positive energy will  worsen any instability problems seen in this simple case.

To begin, we will examine the masses in the supersymmetric AdS vacuum, and ask whether these masses can be made positive definite and large
compared to $|W|$.\footnote{We assume throughout that $F\lesssim \text{few} \times |W|$.} If they can, then after a rigid uplifting to de Sitter space (in the sense described above), the mass matrix will be dominated by ${\cal H}_{\rm susy}$, which is positive definite.
%%v2 few to \text{few}

The dependence of ${\cal{A}}_{i}$ on the $\zeta_a$ can be neglected self-consistently for $Z_{ab}$, but since the nonperturbative contributions lead to an imperfect separability of the superpotential, mass mixings through terms of the form $Z_{ai} $
%%v2 1/10 can not
cannot be neglected, and   the scale of the entries $Z_{ai}$ and $Z_{ij}$ is now  $|W|$. Specifically,
\bea
Z_{ij} &\equiv& {\cal D}_{i} {\cal D}_{j} W = \partial_{i}\partial_{j} W + \left(K_{ij} - K_{i} K_{j}\right) W \,, \\
Z_{aj} &\equiv& {\cal D}_{a} {\cal D}_{j} W = \partial_{a}\partial_{j} W - K_{j} K_{a} W \,,
\eea
where we have used the $F$-flatness conditions $D_i W = D_a W =0$.  The derivatives of $K$ will not lead to enhancements in a controllable regime, while from (\ref{kkltw}) it follows that $\partial_{i}\partial_{j} W \sim \partial_{a}\partial_{j} W \sim W$.  %Thus, for a superpotential of the form (\ref{kkltw}), the supersymmetric masses for the K\"ahler moduli are of order $W$.

The entries of ${\cal H}_{\rm susy}$ %in ${\cal H}_{\rm susy}^{a\bar b}$
 with both indices in the complex structure directions are of order $m^2_{\rm flux} \gg |W|^2$, while the mixed entries in ${\cal H}_{\rm susy}$ receive contributions of order $m_{\rm flux}\ |W|$ from
terms of the form $Z_{a}^{~\bc} \bZ_{\bar j \bar c}$.
The entries in the K\"ahler moduli directions are  of order $|W|^2$.
 %Analogously  to
As in the discussion in \S\ref{sec:decoupling},
the eigenvalues of ${\cal H}$ split into two groups: the first consisting predominantly of
the complex structure moduli and axiodilaton, which are stabilized at a high scale without BF-allowed tachyons, and the second consisting predominantly of the K\"ahler moduli, which have masses of order $|W|$.
Since the supersymmetric K\"ahler moduli masses are not parametrically larger than the negative shift term ${\cal H}_{\rm shift}$, or the off-diagonal contribution ${\cal H}_{\rm pure}$, BF-allowed tachyons are typically abundant in the K\"ahler moduli sector in the supersymmetric AdS vacuum.

Assuming a rigid uplifting to a de Sitter critical point, we recognize that the  K\"ahler moduli sector constitutes a particular variant  of the analysis of \S\ref{sec:MMW} with $F=0$:
the somewhat more favorable regime described in \S\ref{sec:DD}, which requires $|W| \ll |Z_{ij}|$, is inaccessible.  Notice that in the generic regime of \S\ref{sec:MMW}, the Goldstino is by no means the only tachyon, so that instabilities will
arise in the K\"ahler moduli sector even if the  Goldstino direction belongs to some other sector, e.g.\ a local dynamical supersymmetry breaking sector that engineers the positive vacuum energy.

We conclude that if a system described by a superpotential of the form (\ref{kkltw}) is rigidly uplifted to positive vacuum energy, the fraction of de Sitter critical points that are metastable vacua is proportional to
${\rm exp}(-c\, N_K^p)$, with $p > 1$.  For compactifications in which $N_K \equiv h^{1,1}_{(+)}$ is not large, this is not a serious constraint, but it has significant impact for $h^{1,1}_{(+)} \gg 1$, and particularly for  $h^{1,1}_{(+)} \gg h^{2,1}$.

One might object at this point that the arguments in favor of the existence of an approximately-supersymmetric regime \cite{DD}, as in \S\ref{sec:DD}, should hold for general functions $W$, $K$, so why are they not applicable here?  The answer is simply that a superpotential of the form (\ref{kkltw}), which contains one single-instanton term for each K\"ahler modulus, is not a sufficiently general function.
An obvious extension is to consider multiple terms (i.e., a racetrack) for each of the $T_{i}$.  For the purposes of this discussion, we grant any topological prerequisites
for such a multiple racetrack, e.g.\ we suppose that the compactification admits more than one stack of D7-branes in each homology class.  Then, by fine-tuning the fluxes to adjust the prefactors  ${\cal{A}}_{i}$, cf.\ \cite{KL}, one can plausibly arrange that the diagonal entries of $Z_{ij}$ are large compared to $|W|$.  However, there is a statistical price for this fine-tuning, of order $(|W|/m_{\rm susy})^{N_K}$.  Recalling that the boundary between the regimes of   \S\ref{sec:MMW} and \S\ref{sec:DD} occurs for $|W|/m_{\rm susy}\sim 1/N_K$, this fine-tuning is of order  $N_K^{-N_K}$, which can be significant.

To recap, if one assumes a rigid uplifting that changes the cosmological constant without changing the moduli mass matrix, then for a superpotential  of the form (\ref{kkltw}), the  K\"ahler moduli sector will have supersymmetric masses of order $|W|$, and will be governed by the instability analysis of \S\ref{sec:MMW} (with $F=0$, $ W \neq 0$), with positivity probability $P \propto {\rm exp}(-c\, N_K^p)$, with $p>1$.  By fine-tuning a superpotential involving ${\cal O}(N_K)$ racetracks, requiring a statistical price $\sim N_K^{-N_K}$, one can make the supersymmetric masses large enough to guarantee stability.

The situation is considerably worse if supersymmetry is spontaneously broken by an F-term in the K\"ahler moduli sector: again the analysis of \S\ref{sec:MMW} applies generically, but even after fine-tuning  ${\cal O}(N_K)$ racetracks as above, the Goldstino instability will still fall in the K\"ahler moduli sector, so that the instability analysis of \S\ref{sec:DD}, with $p \approx 1.3$, is applicable.

In summary, instabilities appear generic in the K\"ahler moduli sector after uplifting, with metastable de Sitter vacua constituting a fraction $\lesssim {\rm exp}(-c\, N_K)$ of all  de Sitter critical points.
However, for $ h^{2,1} \gg h^{1,1}_{(+)}$, the number of KKLT vacua remains astronomically large, and the overall status of the model is not altered by our findings.

%==============================================================
\section{Conclusions} \label{sec:conclusions}
%==============================================================

We have considered a general four-dimensional ${\cal{N}}=1$ supergravity theory whose superpotential and K\"ahler potential are random functions of $N \gg 1$ scalar fields, and asked what fraction $f$ of de Sitter critical points, with supersymmetry spontaneously broken by an F-term, are metastable vacua rather than unstable saddle points.  Our conclusion is that an exponentially small fraction of critical points are vacua: $f \propto {\rm exp}(-c N^p)$, with $p \gtrsim 1.3$, which differs significantly from earlier results implying $f \sim \frac{1}{N}$.

The character of the instabilities that arise depends on the relative sizes of the supersymmetric and supersymmetry-breaking masses.
At a generic critical point, the soft masses are comparable to the supersymmetric masses, and supersymmetry provides limited protection from instabilities.
We developed a random matrix model for the Hessian matrix ${\cal H}$ at a generic critical point and obtained an analytic formula for its eigenvalue spectrum, finding that a significant fraction of the eigenvalues of ${\cal H}$
are negative.  Eigenvalue repulsion makes large fluctuations of the spectrum statistically costly, and by building on the theory of fluctuations of extreme eigenvalues ---
and through extensive simulations of the full Hessian matrix --- we argued that the probability $P$ of a large fluctuation rendering ${\cal H}$ positive definite is $P \propto {\rm exp}(-c N^p)$, with $p \approx 1.5$ and $c \approx 0.3$.

Eigenvalue repulsion also controls the stability properties of approximately-supersymmetric critical points, at which the F-term $F$ is small compared to the supersymmetric mass scale $m_{\rm susy}$.
In this regime,  only the two eigenvalues corresponding to the Goldstino direction risk becoming tachyonic.
We computed the two smallest eigenvalues to quadratic order in $F/m_{\rm susy}$, and showed that mixing with the  supersymmetric masses shifts these lowest eigenvalues to negative values.
We then studied the probability of a fluctuation to positivity, through analysis of the corresponding univariate statistical distribution and through simulations of the full mass matrix.
In the approximately-supersymmetric regime we found $P \propto {\rm exp}(-c N^p)$, with $p \approx 1.3$ and $c \approx 0.1$.

We emphasize that the assumption that $W$ and $K$ are random functions --- and in particular that their derivatives
are independent random variables drawn from some statistical distribution --- is essential.  There are, however, physically-motivated situations in which $W$ and $K$ are {{\it not}} general random functions of all of their arguments.  An important example consists of two decoupled sectors: if $N_H$ heavy scalars receive large supersymmetric masses, and supersymmetry is dynamically broken in a decoupled system of $N_L$ lighter scalars at a much lower scale, then for a single vacuum configuration of the light fields the corresponding number of vacua of the full system can be exponential in $N_H$.  It seems reasonable to expect decoupling of this sort, into a `degeneracy sector' at high scales, and a dynamical supersymmetry breaking sector at low scales, in a variety of compactifications. Restricting to the $N_L$ light fields, our analysis suggests that the fraction of critical points that are metastable is proportional to  ${\rm exp}(-c {N^p_L})$.  For $N_H \gg N_L$, the result is a mild reduction in the number of metastable vacua.

Let us reiterate: our finding that an exponentially small fraction of critical points in a generic supergravity theory are metastable vacua in no way excludes the existence of a tremendously large landscape of vacua.  There are two primary reasons, one conceptual and one quantitative.
The conceptual reason is the possibility explained in \S\ref{sec:decoupling} and reviewed above of a decoupled system (violating our assumptions on $W$, $K$) in which the vacuum degeneracy is ensured by $N_H$ fields that  receive large supersymmetric masses.  The quantitative reason is that the values of $c, p$ that we have obtained are not so large as to entirely overwhelm the vast number of critical points in flux compactifications.

The methods and results of this work could be of use in understanding the statistical properties of the moduli mass spectrum in string compactifications, and in guiding the search for de Sitter vacua.
One clear implication of our findings is that a direct search for explicit de Sitter vacua in systems with ${\cal O}(10)$ or more fields and reasonably general $W$ and $K$ is likely to be frustrated by the appearance of tachyons.  Correspondingly, the most promising regimes are those in which our assumptions are strongly violated, e.g.\ approximately-supersymmetric critical points for which the superpartner of the Goldstino enjoys special couplings to the remaining fields.  Understanding the incidence of such couplings in well-motivated supergravity theories,  particularly those derived from string compactifications, is an important problem for the future.

\subsection*{Acknowledgements}
We thank J.~Conlon, B.~Czech, S.~Gandhi, B.~Heidenreich, P.~McGuirk, D.~Mehta, E.~Pajer, J.~Sethna, and E.~Silverstein for helpful discussions, and we are grateful to M.~A.~Stephens for correspondence.
We thank M.~Berg, F.~Denef, M.~Douglas, and S.~Kachru for comments on a draft.
L.~M. particularly thanks R.~Easther and S.~Kachru for valuable discussions of this topic over several years.  We thank X.~Chen, G.~Shiu, Y.~Sumitomo, and H.~Tye for informing us of their related work \cite{Chen:2011ac} prior to publication.
This research was supported by the NSF under grant PHY-0757868 and by the Alfred P. Sloan Foundation.
D.~M. gratefully acknowledges the support of the Gålö foundation.
T.~W. was supported by a Research Fellowship (Grant number WR 166/1-1) of the German Research Foundation (DFG).

\appendix

\section{The distribution of critical points} \label{whatisgeneric}
In this appendix we briefly review some pertinent results from the the study of  the distribution of non-supersymmetric vacua  by Denef and Douglas \cite{DD}. In particular, we will review how, for any fixed cosmological constant,   the density of \emph{critical points} --- stable and unstable --- grows linearly with $F$ towards the boundary
of the approximately-supersymmetric regime. This gives evidence for the  expectation that a   ``generic critical point''  typically does not exhibit any particular  hierarchy between $m_{\rm susy}$ and $F$, i.e.~typical critical points in random supergravity are not predominantly of the approximately supersymmetric kind.
  We also comment on how the distribution of metastable vacua is modified by  the exponential suppression of the probability density found in \S\ref{sec:DD}.

The density of critical points  with  cosmological constant $\langle V\rangle = v$ can be evaluated from
\be
N_{\rm crit.pts.}(v) = \int d\mu[W, F, Z, U]\  \delta^{2N}\big(\partial V \big)\  \big| {\rm det} {\cal H} \big|\ \ \delta(V - v) \, . \label{eq:dist}
\ee
Just as in reference \cite{DD}, we assume flat prior probabilities for $W$, $F$, $Z$ and $U$ between $0$ and a cutoff $\Lambda$.  Although this assumption is made here for simplicity, interesting domains of the string theory landscape have been argued to be well-described by these priors.  The measure used in this appendix is
\be
d\mu[W, F, Z, U] = {\cal C}\ d^2W\ d^{2N} F\ d^{k_Z}Z\ d^{k_U}U\ \, ,
\ee
where $k_Z = N(N+1)$, $k_U = \frac{N}{3}(N+1)(N+2)$, and ${\cal C}$ is a normalization constant.

As reviewed in \S\ref{sec:cp}, the critical point equation can be written as an eigenvalue equation for the matrix ${\cal M}$ of equation \eqref{eq:M}, enforcing that ${\cal M}$ has an eigenvalue equal to $2 |W|$ with $\hF$ being proportional to the corresponding eigenvector. Thus, the critical point equation can be simplified by expressing the integration over $Z_{ab}$ (and thereby ${\cal M}$) as an integral over the ordered eigenvalues, $\lambda_1 \leq \ldots \leq \lambda_N$, and unitary rotations ${\cal U}$. Let us denote an orthonormal eigenbasis of ${\cal M}$ as $e^{\pm}_a$, with corresponding  eigenvalues $\pm \lambda_a$. In this basis $\hF$ has components  $|F| \eta^{\pm}_a$, with $\sum_{a=1}^N \Big( (\eta^{+}_a)^* \eta^{+}_a  + (\eta^{-}_a)^* \eta^{-}_a \Big)= 1$. In this notation,
\bea
 \delta^{2N}\big(\partial V \big) &=&  \delta^{2N}\Big( (M-2 |W| ) \hF \Big) = \frac{1}{|F|^{2N}}\ \prod_{a=1}^N\ \delta\Big( \eta^+_a (\lambda_a - 2|W|) \Big)\ \delta\Big( \eta^-_a (-\lambda_a - 2 |W|) \Big) \nonumber \\
 &=& \frac{1}{|F|^{2N}}\ \prod_{a=1}^N\ \delta\Big( \eta^+_a (\lambda_a - 2 |W|) \Big)\ \frac{\delta(\eta^-_a)}{\lambda_a + 2|W|} \, .
\eea
 The cosmological constant constraint $\delta(V-v)$ can be written as
\be
\delta(V-v) = \delta( F^2 - 3 |W|^2 - v ) = \frac{\delta(|W| - w)}{3(|W|+w)} \, ,
\ee
where $w^2 = \frac{1}{3} \left( F^2 - v\right)$. Although the integral \eqref{eq:dist} can also be estimated in the generic regime in which $F \sim m_{\rm susy}$,  this evaluation is slightly technical, and for the purpose of this appendix  it  suffices to discuss  the approximately-supersymmetric regime of Denef and Douglas \cite{DD}. In this case, the integration over $F$ is cut off before $F = m_{\rm susy}/N$, and the  determinant of the Hessian appearing in the integrand of  equation \eqref{eq:dist} is well-approximated by
\be
|{\rm det}{\cal H}| \approx m^2_+\ m^2_-\ \prod_{a'=2}^N \left( \lambda^2_{a'} \right)^2 \, ,\label{eq:detH}
\ee
with $m^2_{\pm}$ as in equation \eqref{eq:Meig}. The number of critical points is
%%v2 1/10 added \times \times
\bea
&&N_{\rm crit.pts.}(v) =
   {\cal C}\ \int d\mu[\vartheta_F, \Omega_F, U, {\cal U}] \int_0^{\Lambda^2} d|W|^2 \int_0^{\epsilon\ m_{\rm susy} } \frac{dF}{F}\    \frac{\delta(|W| - w)}{3(|W|+w)} \times
  \nonumber \\
    && \times \Biggl[ \prod_{a=1}^N \int_{\lambda_{a-1}}^{\lambda_{a+1}} d \lambda_a\  \delta\Big( \eta^+_a (\lambda_a - 2 |W|) \Big)\ \frac{\delta(\eta^-_a)}{\lambda_a + 2|W|} \Biggr]
     f(\lambda_1, \ldots,\lambda_N)\ | \det {\cal H} |\ \,,
\eea
where $f(\lambda_1, \ldots,\lambda_N) $ denotes the joint probability density of equation \eqref{eq:CI},  $\epsilon$ is a small number, and for notational convenience we have defined $\lambda_0 =0$ and $\lambda_{N+1} = \Lambda$. In the approximately-supersymmetric regime, only $\lambda_1$ has a non-negligible probability density at $2 |W|$. With this observation, the integral simplifies to
\bea
&&N_{\rm crit.pts.}(v) =  {\cal C}\ \int d\mu[\vartheta_F, \Omega_F, U, {\cal U}]  \Big(\prod_{a'=2}^N  \delta^{N}(\eta^{+}_{a'}) \Big) \Big(\prod_{a=1}^N  \delta^{N}(\eta^{-}_a) \Big)  \int_0^{\epsilon' \lambda_2 } \frac{dF}{F} \frac{w}{6 w} \times \nonumber \\
   && \times \int_0^{\lambda_2} d\lambda_1 \frac{\delta(\lambda_1 - 2w)}{\lambda_1 + 2 w}
     \left[ \prod_{a'=2}^N \int_{\lambda_{a'-1}}^{\lambda_{a'+1}} d \lambda_a\ \frac{1}{\lambda^2_{a'} - 4w^2} \right]
    f(\lambda_1 = 2 w, \ldots,\lambda_N)\ | \det {\cal H} |\, ,
\eea
where now the constant
%%v2  $\epsilon' < \frac{1}{N}$
$\epsilon' < 1$ encodes the assumed hierarchy between $\lambda_{a'}$ and $F$. Since the probability density  of the smallest eigenvalue exhibits a linear  cleft  for small arguments, cf. equation \eqref{eq:Cleft}, we can (heuristically) write $ f(\lambda_1 = 2 w, \ldots,\lambda_N) = 2 k w\ \tilde f(\lambda_2, \ldots,\lambda_N)$, where $k$ is an ${\cal O}(1)$ constant. This simplifies the integral to
%%v2 = to approx
%%v2 1/10 added \times \times
\bea
&&N_{\rm crit.pts.}(v) \approx
    \frac{k\ {\cal C}}{12}\ \int d\mu[\vartheta_F, \Omega_F, U, {\cal U}]\   \Big(\prod_{a'=2}^N  \delta^{N}(\eta^{+}_{a'}) \Big)\  \Big(\prod_{a=1}^N  \delta^{N}(\eta^{-}_a) \Big) \times \nonumber \\
     && \times \int_0^{\epsilon' \lambda_2 } \frac{dF}{F}\    m^2_+\ m^2_- \
    \left[ \prod_{a'=2}^N \int_{\lambda_{a'-1}}^{\lambda_{a'+1}} d \lambda_a\  \right]
\tilde f(\lambda_2, \ldots,\lambda_N)\  |\det {\cal H'}|^{1/2} \, ,
\eea
where ${\cal H'}$ denotes the truncation of ${\cal H}$ to exclude the Goldstino direction. The scaling of the number of critical points with $F$ at fixed cosmological constant  is evidently determined by the factor
\be
\int_0^{\epsilon' \lambda_2 } \frac{dF}{F}\    m^2_+\ m^2_- \, . \label{eq:AppResult}
\ee
For typical values of $U_{111} \sim m_{\rm susy} $, the Goldstino
%%v2 masses
masses-squared
$m^2_{\pm}$ are each of order $F$, and the number of critical points scales with $F$ as
\be
N_{\rm crit.pts.}(v)  \sim  \int_0^{\epsilon\ m_{\rm susy} } dF\ F \, ,
\ee
and thus, for any given scale of the supersymmetric masses, the critical points are more numerous towards the upper edge of the domain of approximate supersymmetry.

We conclude with some simple remarks. This scaling of the number of critical points with $F$ is consistent with the computation by Denef and Douglas of the scaling of metastable vacua with $F$,
\bea
N_{\rm vacua} \sim \int_0^{\epsilon\ m_{\rm susy}} F^5\ dF \, , \label{eq:Fvacua}
\eea
which is not surprising since the above computation closely mimics that of \cite{DD}. The different scalings of the number of critical points and the number of vacua can be understood from the additional fine-tuning necessary to obtain stability. In the approximately-supersymmetric regime it is necessary to tune $|U_{111}| \lesssim {\cal O}(F)$, which gives an additional factor of $F^2$ from the measure $d|U_{111}|\ |U_{111}|$. Furthermore, as reviewed in \S\ref{sec:DD}, the intent of this fine-tuning is to lower the scale of the Goldstino
mass-squared
%%v2 mass
to ${\cal O}(F^2)$ in order to improve the probability of positivity of ${\cal H}$. By equation \eqref{eq:AppResult}, this provides two more powers of $F$, from which equation \eqref{eq:Fvacua} follows.

Finally, with these flat priors   the additional $N$-dependent (but $F$-independent)  fine-tuning  explored in this paper modifies the density of non-supersymmetric vacua in the approximately supersymmetric regime by
 \be
\frac{
\prod_{a'=2}^{ N^\alpha}\
\int_{0}^{\frac{m_{\rm susy}}{N} }\  d|U_{11a'}|\  |U_{11a'}|   }{
\prod_{a'=2}^{N^\alpha}\ \int_{0}^{\frac{m_{\rm susy}}{\sqrt{N}}} \  d|U_{11a'}|\  |U_{11a'}|} \sim e^{- N^\alpha \ln N} \, ,
\ee
where $N^\alpha$, with $\alpha \le 1$, parameterizes the number of terms in ${\cal S}$ of equation \eqref{sis} that need to be fine-tuned in order for a fluctuation to positivity of the smallest eigenvalue to become likely.

%%%%%%%%%%%%%%%%%%%%%%%%%%%%%%%%%%%%%%%%%%%%%%%%%%%%%%%%%%%%%%%%%%%%%%


\begin{thebibliography}{9}


%\cite{Douglas:2006es}
\bibitem{Douglas:2006es}
  M.~R.~Douglas and S.~Kachru,
  ``Flux compactification,''
  Rev.\ Mod.\ Phys.\  {\bf 79}, 733-796 (2007)
  [hep-th/0610102].


%\cite{DD} {\bf
\bibitem{DD}
  F.~Denef and M.~R.~Douglas,
  ``Distributions of nonsupersymmetric flux vacua,''
  JHEP {\bf 0503}, 061 (2005)
  [hep-th/0411183].


\bibitem{TW} C.~Tracy and H.~Widom, ``Level-spacing distributions and the
Airy kernel," Commun. Math. Phys. {\bf 159}, 151-174, 1994.



\bibitem{Dean2} D.~S.~Dean and S.~N.~Majumdar, ``Large Deviations of Extreme Eigenvalues of Random Matrices," 	Phys.Rev.Lett. {\bf 97} (2006) 160201, arXiv:cond-mat/0609651v2 [cond-mat.stat-mech].

\bibitem{Dean} D.~S.~Dean and S.~N.~Majumdar, ``Extreme Value Statistics of Eigenvalues of Gaussian Random Matrices," 	Phys. Rev. E {\bf 77}, 041108 (2008), arXiv:0801.1730v1 [cond-mat.stat-mech].



\bibitem{Aazami:2005jf}
A.~Aazami and R.~Easther,
  ``Cosmology from random multifield potentials,''
  JCAP {\bf 0603}, 013 (2006)
  [hep-th/0512050].



%\cite{arXiv:0804.1073}
\bibitem{Covi}
  L.~Covi, M.~Gomez-Reino, C.~Gross, J.~Louis, G.~A.~Palma, and C.~A.~Scrucca,
  ``de Sitter vacua in no-scale supergravities and Calabi-Yau string models,''
  JHEP\ {\bf 0806}, 057  (2008)
  [arXiv:0804.1073 [hep-th]].
  %%CITATION = JHEPA,0806,057;%%



\bibitem{free} D.~V.~Voiculescu, K.~J.~Dykema, and A.~Nica, {\it{Free random variables,}} CRM Monograph Series {\bf 1}, American Mathematical Society, Providence, RI, 1992.



%\cite{arXiv:1107.4575}
\bibitem{Zelditch}
  F.~Ferrari, S.~Klevtsov, and S.~Zelditch,
  ``Random Kahler Metrics,''
  arXiv:1107.4575 [hep-th].
  %%CITATION = ARXIV:1107.4575;%%

\bibitem{Rao} A.~Edelman and N.~R.~Rao, ``Random matrix theory,"
Acta Numerica, pp. 1-65, 2005.


\bibitem{Mehta} M.~L.~Mehta, {\it{Random Matrices}}, Academic Press, Boston, 1991.




\bibitem{Wigner0} E.~P.~Wigner, ``On the statistical distribution of the widths and spacings
of nuclear resonance levels," Math. Proc. Cambridge Philos. Soc. {\bf 47}, 790 (1951).

\bibitem{Wigner} E.~P.~Wigner, ``Characteristic vectors of bordered matrices with infinite
dimensions," Annals of Mathematics {\bf 62}, 548-564, 1955.

\bibitem{Wigner1} E.~P.~Wigner, ``Results and theory of resonance absorbtion,"
Oak Ridge Natl. Lab. Rept. ORNL-{\bf 2309}, 1957, 59--70;\\
``Statistical properties of real symmetric matrices with many dimensions,"
Can. Math. Congr. Proc., University of Toronto Press, Toronto, Canada, 1957, 174--184.

\bibitem{Dyson1} F.~J.~Dyson, ``Statistical theory of the energy levels of complex systems, I,'' J.~Math. Phys. {\bf 3} 140 (1962);
``A Brownian motion model for the eigenvalues of a random matrix,'' J. Math. Phys. {\bf 3} 1191 (1962).
%%v2
%\bibitem{Dyson2} F.~J.~Dyson,

\bibitem{Wishartpaper} J.~Wishart, ``The generalized product moment distribution in samples from a normal multivariate population,'' Biometrika {\bf 20A}, 32 (1928).




  \bibitem{Edelman}
A.~Edelman, ``The Distribution and Moments of the Smallest Eigenvalue of a Random Matrix of Wishart Type,''
Linear Algebra and Its Applications, {\bf 	159}, 55, (1991).

\bibitem{MP} V.~A.~Marchenko and L.~A.~Pastur, ``Distributions of eigenvalues of some
sets of random matrices," Math. USSR-Sb {\bf 1}, 507-536, 1967.


\bibitem{AZ}
  A.~Altland and M.~R.~Zirnbauer,
  ``Nonstandard symmetry classes in mesoscopic normal-superconducting hybrid structures,''
  Phys. Rev. B {\bf 55}, 1142 - 1161 (1997).


\bibitem{Forrester} P.~J.~Forrester, ``The spectrum edge of random matrix ensembles," Nucl.
Phys. B {\bf 402}, 709-728 (1993).

\bibitem{Nadal}
C.~Nadal, S.~N.~Majumdar, ``A simple derivation of the Tracy-Widom distribution of the maximal eigenvalue of a Gaussian unitary random matrix", J. Stat. Mech. (2011) P04001, arXiv:1102.0738v3 [cond-mat.stat-mech].




\bibitem{Johnstone}
% computes largest eigenvalues in real Wishart
I.~M.~Johnstone, ``On the distribution of the largest eigenvalue in principal components
analysis,'' Ann. Statist. {\bf 29} (2001), no. 2, 295--327.

\bibitem{Johansson}
% computes largest eigenvalues in complex Wishart
K.~Johansson, ``Shape fluctuations and random matrices," Comm. Math. Phys. {\bf 209}, 437-476, 2000.


\bibitem{glass} G.~Ben Arous, A.~Dembo, and A.~Guionnet, ``Aging of spherical spin
glasses," Probab. Theory Relat. Fields, {\bf 120}, 1-67 (2001).


\bibitem{Bray}
A.~J.~Bray and D.~S.~Dean, ``The statistics of critical points of Gaussian fields on large-dimensional spaces,'' arXiv:cond-mat/0611023v1 [cond-mat.dis-nn].




\bibitem{Vivo}
P.~Vivo, S.~N.~Majumdar, and O.~Bohigas, ``Large Deviations of the Maximum Eigenvalue in Wishart Random Matrices,''
J. Phys. A: Math. Theor. {\bf 40} (16) (2007) 4317-4337, arXiv:cond-mat/0701371v2 [cond-mat.stat-mech].



\bibitem{Vergassola} S.~N.~Majumdar and M.~Vergassola, ``Large Deviations of the Maximum
Eigenvalue for Wishart and Gaussian Random Matrices,'' Phys. Rev. Lett. {\bf 102}, 060601 (2009), arXiv:0811.2290.

\bibitem{Nadal2}
S.~N.~Majumdar, C.~Nadal, A.~Scardicchio, and P.~Vivo, ``The Index Distribution of Gaussian Random Matrices,'' Phys. Rev. Lett. {\bf 103}, 220603 (2009), arXiv:0910.0775v1 [cond-mat.stat-mech].





\bibitem{KC} E.~Katzav and I.~P.~Castillo, ``Large Deviations of the Smallest Eigenvalue of the Wishart-Laguerre Ensemble,'' arXiv:1005.5058v3 [cond-mat.dis-nn].



\bibitem{Speicher}
R.~Speicher,
``Multiplicative functions on the lattice of non-crossing partitions and free convolution,'' Math. Ann. {\bf 298} (1994), 611-628.





\bibitem{RaoPoly} A.~Edelman and N.~R.~Rao, ``The polynomial method for random matrices," Foundations of Computational Mathematics, arXiv:math/0601389v3 [math.PR].



\bibitem{Abramowitz} M.~Abramowitz and I.~A.~Stegun, ``Handbook of Mathematical Functions with Formulas, Graphs, and Mathematical Tables,'' Dover, New York, 1964.


\bibitem{Solomonson}
H.~Solomon and M.~A.~Stephens, ``Distribution of a Sum of Weighted Chi-Square Variables,''
Journal of the American Statistical Association {\bf 72} 360 (1977), 881-885.
%%v2

\bibitem{lc}
K.~A.~Takeuchi and M.~Sano, ``Universal Fluctuations of Growing Interfaces: Evidence in Turbulent Liquid Crystals,'' Phys. Rev. Lett. {\bf 104}, 230601 (2010).

\bibitem{bus}
J.~Baik, A.~Borodin, P.~Deift, and T.~A.~Suidan, ``A model for the bus system in Cuernavaca
(Mexico),'' J. Phys. A {\bf 39} (28) (2006), 8965 - 8975; M.~Krb\'{a}lek and P.~\v{S}eba, ``Statistical properties of the city transport in Cuernavaca
(Mexico) and random matrix theory,'' J. Phys. A: Math. Gen. {\bf 33} (2000), 229 - 234.


\bibitem{lasers}
M.~Fridman et. al., ``Measuring maximal eigenvalue distribution of Wishart
random matrices with coupled lasers,'' arXiv: 1012.1282.


\bibitem{Deift} P. Deift, ``Universality for mathematical and physical systems,"
arXiv:math-ph/0603038v2.



\bibitem{KUniv}
A.~B.~J.~Kuijlaars, ``Universality,'' Chapter 6 in {{\it Oxford Handbook of Random Matrix Theory}}, G. Akemann, J. Baik, and P. Di Francesco, eds., Oxford University Press, 2011.
arXiv:1103.5922v2 [math-ph].


\bibitem{Bai} Z.~D.~Bai, ``Methodologies in spectral analysis of large dimensional random
matrices, a review," Statistica Sinica {\bf 9}, 611-677.


\bibitem{Soshnikov} A.~Soshnikov, ``A note on universality of the distribution of
the largest eigenvalues in certain sample covariance matrices," J. Statist.
Phys. {\bf 108}, 1033--1056, 2002.



\bibitem{heavy} G.~Biroli, J-P.~Bouchaud, and M.~Potters, ``On the top eigenvalue of heavy-tailed random matrices," cond-mat/0609070.


%\cite{Borghese:2011en}
\bibitem{Borghese:2011en}
  A.~Borghese, R.~Linares and D.~Roest,
  ``Minimal Stability in Maximal Supergravity,''
  arXiv:1112.3939 [hep-th].
  %%CITATION = ARXIV:1112.3939;%%


\bibitem{SSB}
J.~H.~Schenker and H.~Schulz-Baldes, ``Semicircle law and freeness for random matrices with symmetries or correlations,'' Math. Res. Lett. {\bf 12} (2005), 531-542, arXiv:math-ph/0505003v1.


%\cite{hep-th/0307049}
\bibitem{AshD}
  S.~Ashok and M.~R.~Douglas,
  ``Counting flux vacua,''
  JHEP\ {\bf 0401}, 060  (2004)
  [hep-th/0307049].
  %%CITATION = JHEPA,0401,060;%%





\bibitem{DSZ}
M.~R.~Douglas, B.~Shiffman, and S.~Zelditch,
  ``Critical points and supersymmetric vacua, I, II, III,''
  Commun. Math. Phys. {\bf 252} (2004) 325-358, arXiv:math/0402326v2 [math.CV];
% M.~R.~Douglas, B.~Shiffman, and S.~Zelditch,
%  ``Critical points and supersymmetric vacua, III: String/M models''
 % M.~R.~Douglas, B.~Shiffman, and S.~Zelditch,
%  ``Critical points and supersymmetric vacua, II: Asymptotics and extremal metrics,''
  J. Diff. Geometry {\bf 72} (2006), 381-427, arXiv:math/0406089v3 [math.CV];
Commun. Math. Phys. {\bf 265}:617-671, 2006, arXiv:math-ph/0506015v4.


%\cite{hep-th/0301240}
\bibitem{KKLT}
  S.~Kachru, R.~Kallosh, A.~D.~Linde, and S.~P.~Trivedi,
  ``De Sitter vacua in string theory,''
  Phys.\ Rev.\ D\ {\bf 68}, 046005  (2003)
  [hep-th/0301240].
  %%CITATION = PHRVA,D68,046005;%%



%\cite{hep-th/0404257}
\bibitem{DDF}
  F.~Denef, M.~R.~Douglas, and B.~Florea,
  ``Building a better racetrack,''
  JHEP\ {\bf 0406}, 034  (2004)
  [hep-th/0404257].
  %%CITATION = JHEPA,0406,034;%%

%\cite{hep-th/0503124}
\bibitem{DDFGK}
  F.~Denef, M.~R.~Douglas, B.~Florea, A.~Grassi, and S.~Kachru,
  ``Fixing all moduli in a simple F-theory compactification,''
  Adv.\ Theor.\ Math.\ Phys.\ \ {\bf 9}, 861  (2005)
  [hep-th/0503124].
  %%CITATION = 00203,9,861;%%





%\cite{Bousso:2000xa}
\bibitem{Bousso:2000xa}
  R.~Bousso and J.~Polchinski,
  ``Quantization of four form fluxes and dynamical neutralization of the cosmological constant,''
  JHEP {\bf 0006}, 006 (2000)
  [hep-th/0004134].

%\cite{hep-th/0411011}
\bibitem{KL}
  R.~Kallosh and A.~D.~Linde,
  ``Landscape, the scale of SUSY breaking, and inflation,''
  JHEP\ {\bf 0412}, 004  (2004)
  [hep-th/0411011].
  %%CITATION = JHEPA,0412,004;%%

%\cite{Chen:2011ac}
\bibitem{Chen:2011ac}
  X.~Chen, G.~Shiu, Y.~Sumitomo and S.~H.~H.~Tye,
  ``A Global View on The Search for de-Sitter Vacua in (type IIA) String Theory,''
  arXiv:1112.3338 [hep-th].
  %%CITATION = ARXIV:1112.3338;%%


\end{thebibliography}
\end{document}